`

# Photonic Theta Cavity: Engineering Bound States in the Continuum in Topological Resonators Beyond the Limitations of Near-Field Coupling

**ERIC SEABRON[1*], ROBERT E. COLEMAN.[1], CHUANYU LIAN[2], MALCOLM BOGROFF[1], CARLOS RÍOS[2], ANTONIO LEVY[3]**

[1]*Howard University, Department of Electrical and Computer Engineering, Washington, DC 20059, United States*
[2]*University of Maryland, Departments of Materials Science and Engineering, College Park, Maryland 20742, United States*
[3]*Howard University, Department of Physics, Washington, DC 20059, United States*
**eric.seabron@howard.edu*

**Abstract:** The Theta Cavity is a unique topological resonator architecture which utilizes interferometric coupling to overcome fundamental design limitations associated with near-field evanescent coupling which currently dominates the design space for integrated photonics. The defining device physics is established by the mirror symmetric cross junctions which preserves efficient power transfer between a waveguide and ring resonator creating a strongly correlated phase relationship between the multiple paths. The unique mode selection physics allows for interference driven suppression of radiative pathways enabling strong cavity confinement and the emergence of Bound States in the Continuum (BICs). Analytical models reveal non-Hermitian optical band structure displaying non-trivial topological transitions between BIC and quasi-BIC modes that are robust to attenuation, temperature variations, and typical fabrication non-idealities, which is critical for overcoming intrinsic limitations associated with silicon based integrated photonics. The Theta Cavity architecture also circumvents limitations that arise from proximity requirements of the physical "gap" used in near-field coupled waveguides which enables a flexible design space for new devices. In this study, we demonstrate phase-mediated long-range strong coupling of multiple ring resonators in the Nested Theta Cavity architecture showcasing band structure hybridization resulting in the formation of anti-crossing bandgaps, Dirac crossings, and Fano resonances. The photonic Theta Cavity architecture provides a scalable, topologically robust platform for engineering modes across a multi-dimensional parameter space with high resilience to perturbation and attenuation, enabling a new approach for the designing of integrated cavity devices.

## Introduction

Engineering efficient coupling between waveguides and resonators is essential to the design of integrated photonic systems. Conventional inline photonic integrated circuits devices utilize wave interference to enable functionality and strong coupling; notable architectures include Mach-Zender interferometers [1–4], conventional microring and racetrack ring resonators [5–10], Fabry-Perot Cavities [11–15], and Multimode Interferometers [11,16,17]. Near-field coupling is a widely used design convention which takes advantage of overlapping evanescent fields in parallel waveguides to facilitate resonant tunnelling and subsequent energy transfer [18–22]. For example, in conventional ring resonators the near-field coupling region are designed with a low-index, sub-micron (100 nm – 500 nm) "gap" between the waveguide and ring resonator [20,23–25]. Racetrack resonators employ coupling regions with extended lengths (several microns)

`

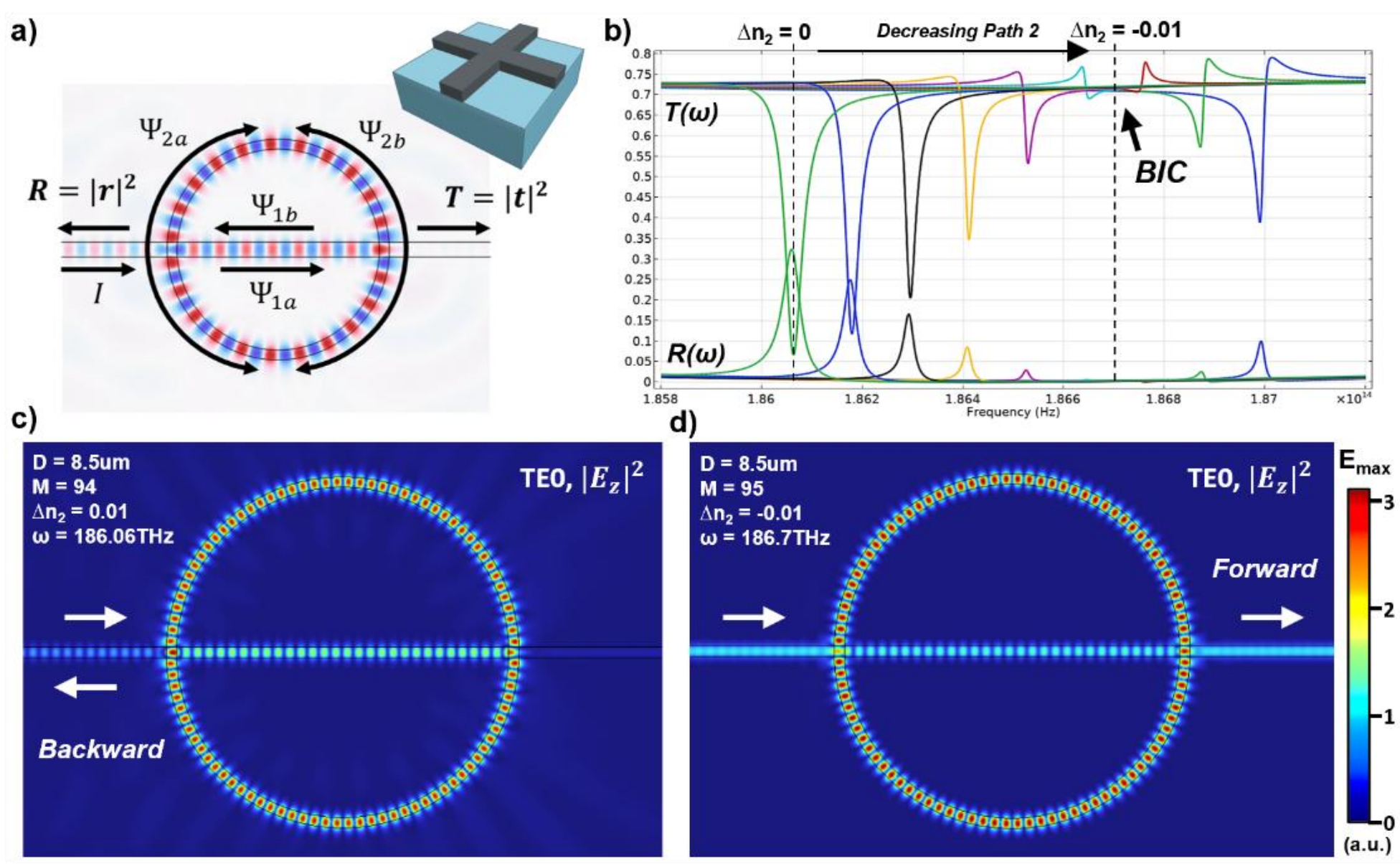

**Fig 1: Theta Cavity Architecture.** (a) Theta Cavity physical model, the simulated electric field illustrates interferometric coupling between the middle waveguide (path 1, $\Psi_1$) and ring resonator (path 2, $\Psi_2$), the inset shows the three-dimensional geometry of the mirror symmetric cross junction (MSCJ). (b) Simulated transmission spectra (upper curves, T(ω)) and reflection spectra (lower curves, R(ω)) for a single mode are overlayed to reveal a redshift and transition in spectral signature with increasing refractive index in the ring resonator (right to left). The simulated normalized Electric Field of a representative Transverse Electric (TE0) Theta Cavity Mode display it's the characteristic standing wave for the forward propagating, M = 94 (c), and backward propagating, M = 95 (d), condition.

to improve the extinction ratio and stability of ring resonator modes [6,9,26–28]. Limitations in conventional architecture arise from fundamental tradeoffs between energy transfer, mode quality factor, fabrication tolerances, and design constraints imposed by engineering near-field coupling regions. The Theta Cavity (TC) employs interferometric principles to facilitate coupling between the waveguide transmission line and the ring resonator. The novelty of the Theta Cavity architecture lies in its unique application of mirror symmetric, "cross" shaped coupling regions, denoted as mirror-symmetric cross (MSC) junctions. This design facilitates interference to control flow between orthogonal paths formed at the intersection of the waveguide "path 1" and ring resonator "path 2" (Fig 1a). The TC mode selection is determined by the phase relation between the two paths that comprise the TC's geometry, resembling the shape of the Greek letter *Theta* Θ. Like conventional ring resonator modes, TC modes are excited when the round-trip path length of the ring resonator approaches an integer multiple of $2\pi$, forming a distinct standing wave within the ring cavity [29,30]. The MSC junction's mirror symmetry ensures a robust phase condition between the clockwise (CW) and counterclockwise (CCW) propagating ring modes which preserves the standing wave formation at the resonant cavity condition. In contrast to parallel junctions, this architecture always maintains highly efficient power transfer between the waveguide and ring resonator, due to the lack of physical barriers, instead mode selection is determined by the cavity geometry eliminating design constraints imposed by near-field coupling. The high finesse of the ring resonator induces a periodic phase condition for constructive and destructive interference between the two paths leading to the emergence of a unique band structure with non-trivial topologically protected transition [31–33]. Overall, the Theta Cavity architecture represents a completely new approach to

`

engineering photonic band structure offering uniquely exotic physics for the advancement of Non-Hermitian Topological Photonics.

A representative two-dimensional electromagnetic simulation model (2D COMSOL – Electromagnetic Waves Frequency Domain) is used throughout this study to qualitatively explore the physics of the TC. In Fig 1b, the simulated transmission and reflection spectra reveal the effects of tuning the index of the ring resonator (path 2) relative to the middle waveguide (path 1). A distinct behavior of the Theta Cavity is the outcoupled mode (Fig 1c), which propagates in the backward direction appearing as a peak in the reflection spectra and dip in the transmission spectra, can destructively interfere to suppress mode leakage through the ports. This interaction manifests as an abrupt change in outcoupled mode propagation from backward to forward direction, indicated by the formation of a Fano line shape observed in the spectra (Fig 1b), showing a rapid transition from enhanced transmission to mode extinction. Further tuning the phase condition results in complete destructive interference of the outcoupled radiation, where the mode signature is hidden despite the presence of the bound state cavity mode revealed in the simulated electric field maps (Fig 1d). Further analysis, validated by simulation models and experimental devices, validates our observation of an emergent Bound State in a Continuum (BIC) transition [34–36]. The inherent symmetry of the Theta Cavity geometry protects the Quasi-BIC to BIC transition from perturbations, while its unique non-Hermitian topology preserves the band structure under modest attenuation. For state-of-the-art platforms, BICs transitions typically degrade to Quasi-BICs (qBICs) with a finite bandwidth with increasing attenuation, which contributes to their limited observation in integrated photonic platforms where light absorption and scattering is prevalent [33,37–41]. Non-intuitively, introducing attenuation in the Theta Cavity reinforces the topologically protected interference expanding the region of propagation inversion, which describes the physical manifestation of the BIC transition. We assert that Theta Cavity BIC transitions represent a new functionality which circumvents conventional limitations requiring ultrahigh quality factor modes to enhance device sensitivity. Another notable advantage of the Theta Cavity is the ability to control the extinction coefficient using two physical degrees of freedom, path 1 and path 2. Conventional ring resonators typically utilize a single degree of freedom, the effective round-trip path length, which shifts the free spectral range. The TC offers new functionality for amplitude modulation at a fixed frequency by tuning the index of the middle path, explored in Fig S1. In this study, we present an analytical framework, validated by simulation models and device measurements, to predict mode selection and demonstrate phase mediated strong coupling with unique non-Hermitian physics not observed in conventional photonic resonator architectures. Altogether, the Theta Cavity architecture offers significant potential for spectral engineering in platforms limited by non-idealities such as lower resolution lithography (imperfect geometries), environmental variability (temperature fluctuations), and lossy material platforms (eg Silicon-On-Insulator).

The physical interpretation of a Theta Cavity BIC is best described as a perfect standing wave cavity mode occurring within a continuum of propagating waveguide modes which freely pass through the cavity. The radiative pathway between the ring resonator and waveguide can be shut off via destructive interference at both ports. Constructive interference at both ports results in a "leaky" cavity mode observed as a broad resonance, and destructive interference in only one of the ports results in a strongly confined cavity mode observed as a sharp resonance with a characteristic backward propagation of outcoupled light (Fig 1c). However, when the phase condition of the cavity supports perfect destructive interference occurs at both ports, the radiative pathways between the cavity mode and the propagating waveguide modes are completely shut off forming a Bound State in the Continuum (Fig 1d) resulting in the complete suppression of an observable resonance in the spectra (Fig 1b). According to the physical model, orthogonality at the cross junctions imposes a boundary condition in which the coupling between path 1 and path 2 is identical for both the clockwise (CW) and counterclockwise

`

(CCW) ring modes. This boundary condition preserves the standing wave condition in both path 1 and path 2 for all TC modes and ensures that perfect destructive interference of the radiative pathways at the ports remains achievable to support the BIC condition. Enforcing mirror symmetry at the cross junction as a design constraint allows the orthogonal components of the CW and CCW ring modes to be consolidated into a single effective wavefunction, represented in the analytical model with a single phase parameter and coupling coefficient. The MSC junction protects the desired TC boundary condition which preserves robust BIC formation independent of ring resonator geometry. In addition, by designing the TC with equal path lengths in the upper and lower parts of path 2, only two wavefunctions ($\Psi_1$ and $\Psi_2$ in Fig 1a) are required to describe interference at each port. These design constraints ensure the phase condition required for perfect destructive interference of the radiative pathways is inherently maintained and predictable.

## Analytical Model

When light interacts within the MSC junctions, the energy propagating in the waveguide (path 1) is transmitted or reflected into the ports or couples to the orthogonal ring resonator (path 2). We assume a steady state power flow through the MSC junctions allowing for a closed set of simultaneous equations to describe the interactions within the TC. The analytical model uses three interface-continuity equations for each port (Eq. 1-2) to solve the wavefunctions along both paths described in the TC physical model (Fig 1a).

$$\begin{cases} t = \Psi_{1b}e^{-i\phi_{1b}} + \Psi_{1a}e^{i\phi_{1a}} \\ t = \Psi_{2a}e^{i\phi_{2a}} + \Psi_{2b}e^{-i\phi_{2b}} \\ t = z_1(\Psi_{1a}e^{i\phi_{1a}} - \Psi_{1b}e^{-i\phi_{1b}}) + z_2(\Psi_{2a}e^{i\phi_{2a}} - \Psi_{2b}e^{-i\phi_{2b}}) \end{cases} \quad \text{Eq. (1)}$$

$$\begin{cases} 1 + r = \Psi_{1a} + \Psi_{1b} \\ 1 + r = \Psi_{2a} + \Psi_{2b} \\ 1 - r = z_1(\Psi_{1a} - \Psi_{1b}) + z_2(\Psi_{2a} - \Psi_{2b}) \end{cases} \quad \text{Eq. (2)}$$

Mirror symmetry is represented in the system of equation by applying the following assumptions $\phi_{1a} = \phi_{1b}, \phi_{2a} = \phi_{2b}$. Design coefficients for effective coupling ($z$) and loss ($\delta$) are introduced to capture tradeoffs and fit the analytical model. The effective coupling coefficient describes the propensity of modes in the waveguide to effect modes in the ring pathways and vice versa. When the coupling between the paths is weaker, the z-coefficient is small, implying stronger confinement of light stored within the ring resonator. For the TC devices we assume $z_1 = 1$ and $z_2 = z < 0.5$, implying the coupling between the path 1 and path 2 is the same for the CW and CCW modes at both ports, with more power flowing in the direction of propagation than in the orthogonal paths within the junction. The loss coefficient ($\delta$) is a unitless constant used to capture all forms of attenuation within the resonator, incorporated as the imaginary part of the phase, such that $\phi \sim \phi + j\delta$. The mode number ($M$) is determined by the number of standing wave nodes at the resonant cavity condition, where the round-trip path length in the ring resonator (path 2) approaches an integer multiple of the wavelength. The geometric tuning parameter ($\gamma$) relates the effective path length of path 1 (center waveguide) to that of path 2 (ring resonator), where $\gamma = L_1/L_2$, allowing the reduction of variables; $\phi_2 = \phi$, $\phi_1 = \gamma\phi$. The phase of middle waveguide path ($\phi_1 \sim \gamma$) at the resonant cavity condition, approximated by $\phi_1 = \gamma M\pi$, is used as a direct analogous to the normalized wavenumber ($k$) to map the Theta Cavity band structure. The characteristic interface-continuity equations are used to derive a closed form equation for the transmission (Eq. 3), reflection (Eq. 4), and electric field enhancement spectra (Eq. 5) determined by the TC design coefficients, discussed in more detail in supplemental section. These equations describe the TC dispersion

`

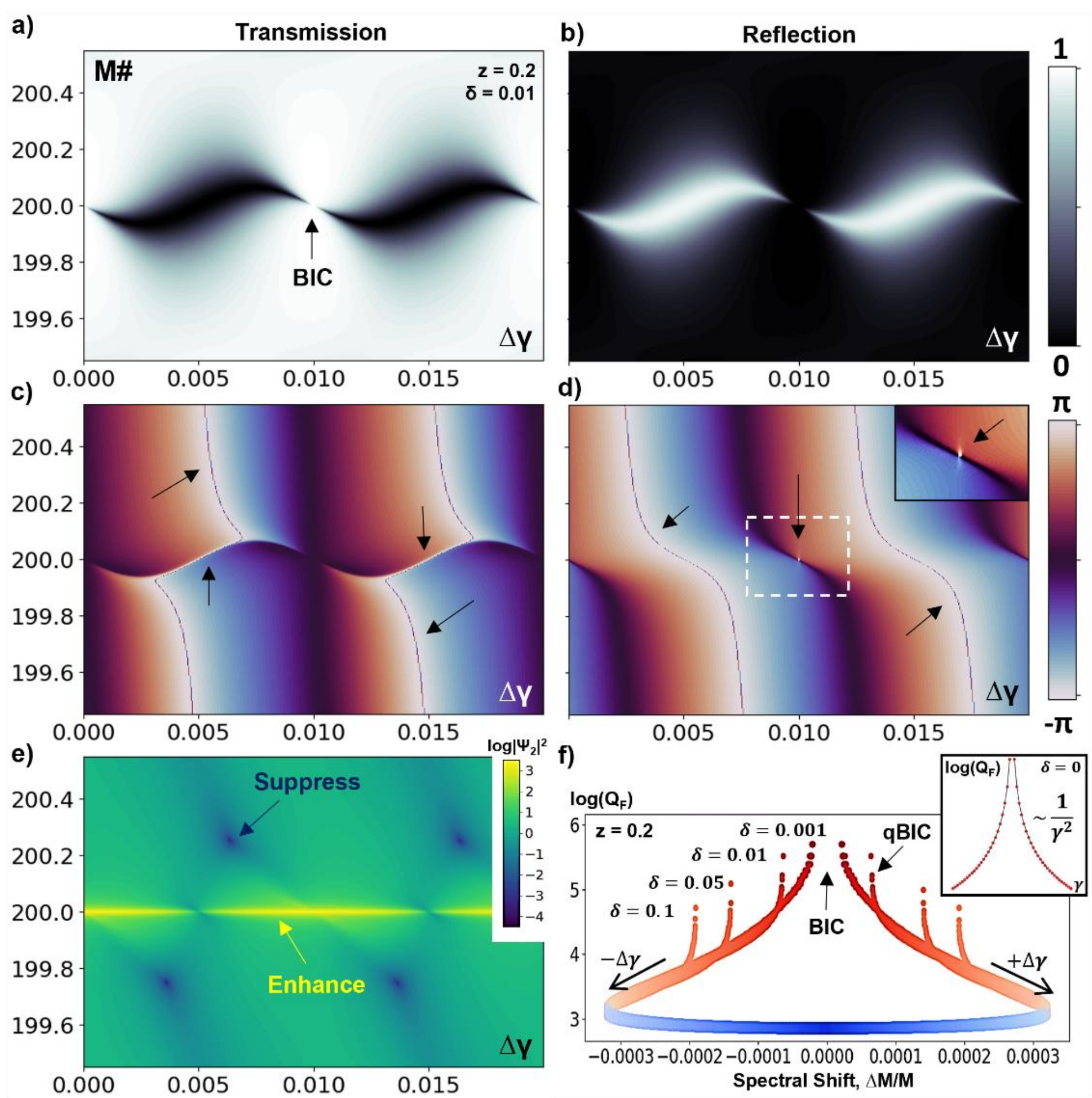

**Fig 2: Topological Bound State in the Continuum.** The normalized dispersion relations are generated numerically from the analytical model, where the x-axis represents the differential gamma (Δγ) and the y-axis is the Mode number (M). The magnitude of the transmitted (a) and reflected (b) wavefunction reveal spectral bands, the arrow highlights propagation inversion. The phase of the transmitted (c) and reflected (d) wavefunction are false colored to reveals discontinuity features, the arrows highlight non-trivial topological band structure. (e) The magnitude of the electric field (log10 scale) confined to the ring resonator $|\Psi_2|$, show regions of Electric field enhancement (false yellow) and suppression (false blue). (f) The Non-Hermitian Phase Manifold (NHPM) predicts the Quality Factor and Spectral Shift of Theta Cavity Modes as the relative phase ($\Delta\phi_1 \sim \Delta\gamma$) varies from 0 to 2π (false color red – grey – blue); trends are overlayed with increasing attenuation (δ) revealing the opening of the manifold, the inset plots the Quality Factor ($\log Q_F$) versus the geometric parameter (γ) under zero attenuation (δ=0) showing a $1/\gamma^2$ trend for the qBIC to BIC transition.

relation without significant computation resources associated with solving partial differential equations and finite element analysis simulations.

$$r(\phi) = \frac{(zsin(\gamma\phi)+sin(\phi))(-2z(cos(\phi)-e^{j\gamma\phi})+jz^2 sin(\phi))+j((zcos(\phi)-ze^{j\gamma\phi})^2-2jzsin(\phi))}{(zsin(\gamma\phi)+sin(\phi))(4zcos(\phi)-j(z^2+4)\,sin(\phi)-2ze^{j\gamma\phi})-j(z(cos(\phi)-e^{j\gamma\phi})-2j\,sin(\phi))^2} \quad \text{Eq. (3)}$$

$$t(\phi) = \frac{2z(zsin(\gamma\phi)+sin(\phi))}{(zsin(\gamma\phi)+sin(\phi))(4zcos(\phi)-j(z^2+4)\,sin(\phi)-2ze^{j\gamma\phi})-j(z(cos(\phi)-e^{j\gamma\phi})-2j\,sin(\phi))^2} \quad \text{Eq. (4)}$$

$$\Psi_2(\phi) = \frac{(1+r)e^{-j\phi}}{2j\,sin(\phi)} - t \quad \text{Eq. (5)}$$

The normalized dispersion relation is revealed by plotting the geometric tuning parameter ($\gamma$) along the x-axis, representing phase relationship of the paths in TC, and mode number (*M*) along the y-axis, proportional to the frequency ($\omega$); $z = 0.2$, $\delta = 0.01$ is assumed for analysis. The dispersion relation for the magnitude of the transmission (Fig 2a) and reflection wavefunction (Fig 2b) reveals a band structure, showing notable spectral shifts and rapid narrowing of the FWHM as the relative phase difference between path 1 and path 2 approaches even multiples of $\pi$. Investigation of the dispersion relation for the phase of the transmission (Fig 2c) and reflection wavefunction (Fig 2d) reveals asymptotic discontinuities and anomalous transitions in outcoupled phase highlighting non-trivial topological transitions [42,43]. The BIC mode singularity, highlighted in the inset of the reflected phase dispersion (Fig 2d), shows an abrupt $\pi$-shift in phase, characteristic of non-Hermitian band structure [42–44]. In Fig 2e, the magnitude of the wavefunction confined to path 2 (ring resonator) reveals strong electric field enhancement for integer mode numbers (*M*) as expected for strongly confined cavity modes. The enhancement dispersion also reveals interesting eigen solutions where the mode coupling is significantly suppressed by destructive interference in the ring resonator, appearing as blue spots in false color plot, which correspond to anomalous transition revealed in the phase dispersion (Fig 2cd). The vanishing transmission and reflection signatures (Figs. 2a–b) converge with strong TC mode confinement (Fig. 2e) as the tuning parameter ($\gamma$) approaches a symmetric phase condition at both MSC junctions, identifying the parametric Bound in the Continuum (BIC) transition [44,45]. The topological nature of Theta Cavity BIC modes is further evidenced by their parametric independence. While variations in the device geometry and the waveguide properties, including attenuation, may shift the resonant frequencies as expected, the MSC junctions ensure the robustness of TC modes and preservation of the BIC transition. In Fig 2f, the Quality Factor ($Q_F$) of the TC modes is plotted against the gamma parameter ($\gamma$) revealing the BIC transitions into a Quasi-BIC with the detuning following a $Q_F \sim \gamma^{-2}$ scaling trend when assuming zero loss ($\delta = 0$), which is a direct consequence of the BIC acting as topological charge defect [44,46,47]. Our analysis is in conclusive agreement with physical observations and interpretations for identifying BIC signatures [33–36,44–48].

The phase-coupling tradeoffs, spectral shift and quality factor versus the tuning parameter, is captured in the Non-Hermitian Phase Manifold (NHPM), shown in Fig 2f. Typically, BICs are fragile to attenuation and perturbations which affect the tuning parameter. However, in the TC the unique qBIC to BIC transition, represented as an asymptotic trend in quality factor coinciding with the opening of the NHPM, is preserved with increasing loss coefficient ($\delta > 0$), ambient of non-Hermitian behavior [49,50]. In addition, the tuning parameter ($\gamma$) is unaffected by perturbations that equally affect all paths simultaneously which improves the stability of Theta Cavity modes. This implies that TC modes are highly sensitive to local perturbations which affect the paths individually while exhibiting robustness against global perturbations. Theta Cavity BICs are parametric in nature, where the predictable BIC condition is not dependent on coupling strength and attenuation due to the imposed symmetric design constraints. The physical description of TC-BICs have notable similarities to Friedrich-Wintgen BICs (FW-BICs) since they arise from destructive interference of the ring resonator (path 2) and

`

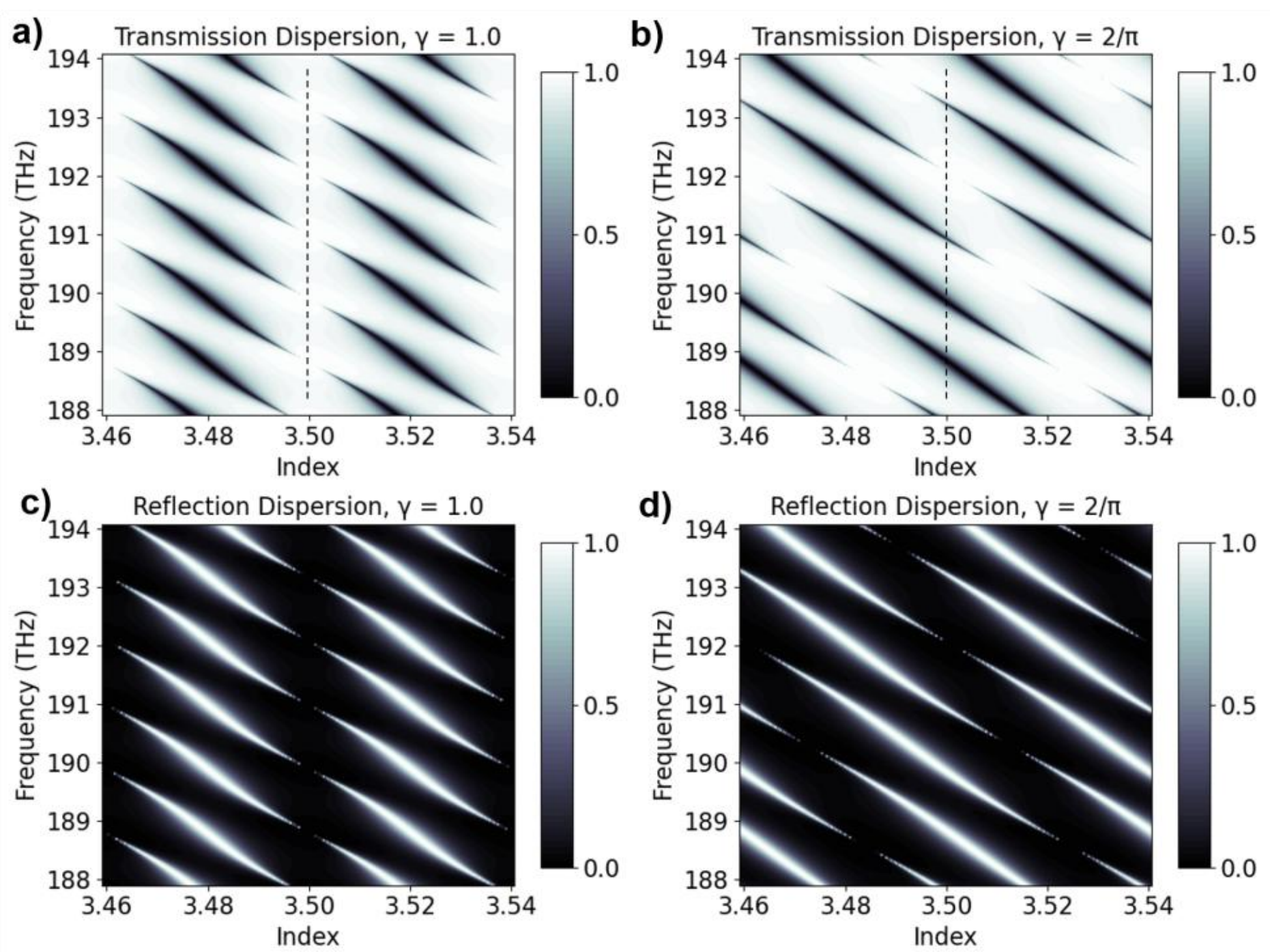

**Fig 3: Theta Cavity Dispersion Relation.** The plots show the analytical dispersion relation TC device, a ring circumference of 100um and effective index of 2.7, the x-axis is the refractive index of the ring resonator (path 2) and the y-axis is frequency (THz). The transmission and reflection dispersion is shown for γ = 1 (a,c) and γ = 2/π (b,d) respectively. The horizontal dashed lines approximate the spectral pattern at a fixed index.

waveguide (path 1) wavefunctions through tuning of the geometric parameter gamma ($\gamma$) [45,51]. However, TC-BICs are also distinct because the ring resonator and waveguide components of the TC mode do not constitute separable resonances. Consequently, unlike conventional FW-BICs, TC-BICs do not exhibit mode splitting under detuning conditions [51]. An important characteristic of the Theta Cavity BIC transition is that ultrahigh quality factor is not required to produce a rapid quasi-BIC to BIC transition, as shown by the NHPM (Fig 2f). This preserves the high contrast event in platforms with modest attenuation (eg Silion-on-Insulator photonic platforms) which is important for developing new functionality.

The device dispersion relation, plotted as frequency of incoming light (THz) versus ring resonator refractive index in Fig 3, can is generated from the normalized dispersion (Fig 2ab) by incorporating device parameters, the effective index and TC geometry. The predicted analytical spectra pattern, for a ring circumference of 100 um, is determined by taking a linecut at a fixed refractive index revealing distinct mode patterns in the spectra determined by the geometric coefficient (γ). When path 1 and path 2 have the same path length (γ = 1), the phase relation is the same for every mode in the spectra creating a repeating mode pattern (Fig 3ac). However, for a circular Theta Cavity (γ = 2/π), the phase shift with increasing M-number is inherently irrational leading to an aperiodic spectral pattern (Fig 3bd). In Fig 4, the analytical dispersion relation model was validated against the electromagnetic simulation model successfully predicting the spectral behavior of the fundamental waveguide modes. Electric field maps of the Transverse Electric, TE0 (Fig 4a), and Transverse Magnetic, TM0 (Fig 4d), TC modes at the MSC junctions reveal distinct "supermodes" formation, which facilitates light coupling between orthogonal paths and leakage into the far field, contributing to differences in their respective z-coupling and δ-attenuation. We successfully fit the analytical spectra to the

`

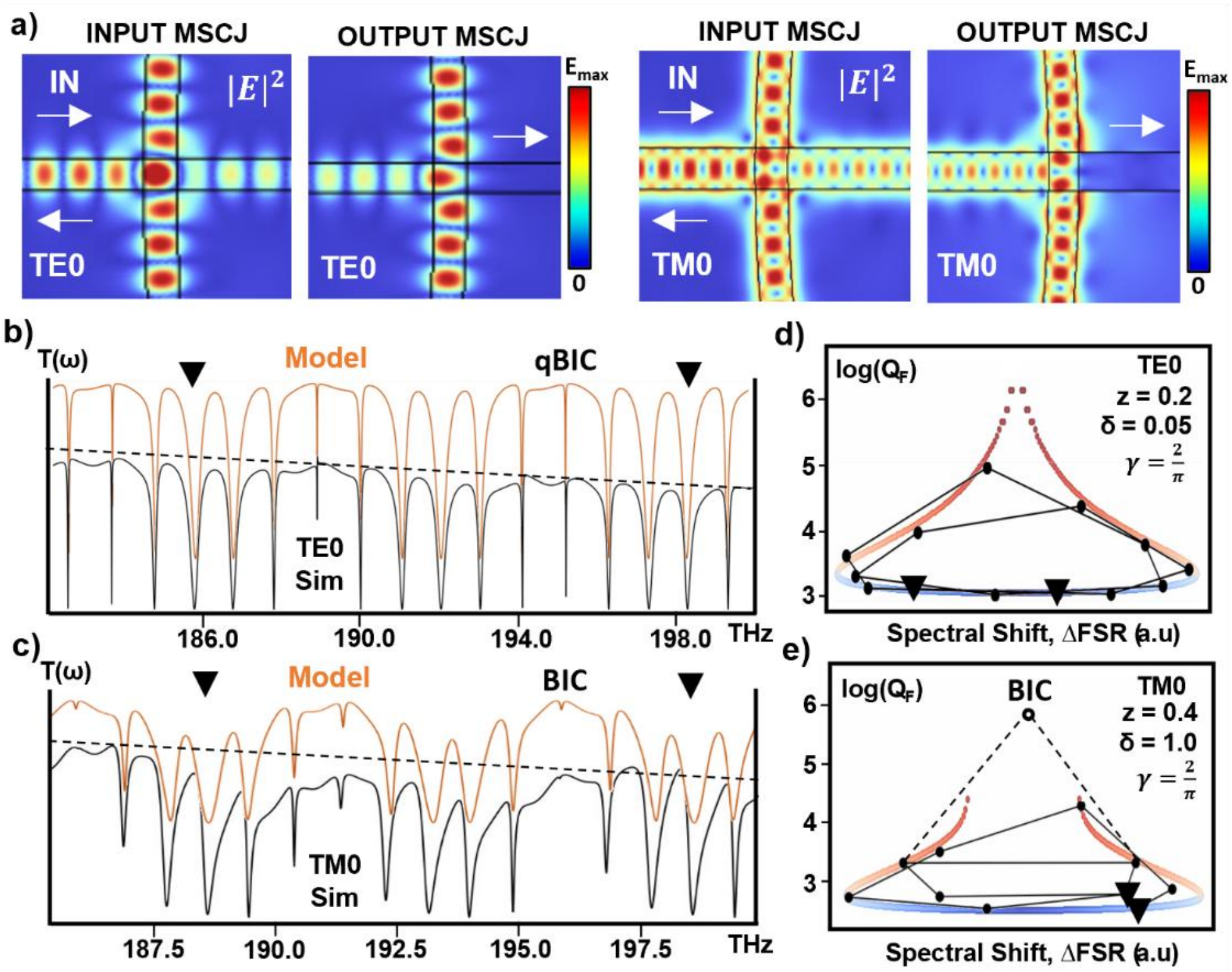

**Fig 4: Fundamental TE0 and TM0 Theta Cavity modes.** (a) The simulated Electric field maps resonant coupling behavior in the input (left) and output (right) mirror symmetric cross junction (MSCJ) for the TE0 mode (top) and TM0 modes (bottom). The transmission spectra predicted by the analytical (lower, black) and simulation model (upper, false color) are overlayed with an artificial offset for the TE0 mode (b) and TM0 mode (c). The Quality Factor (dB) and spectral shift of the modes extracted from simulated TE0 (d) and TM0 (e) spectra is plotted overlayed with the non-Hermitian phase manifold predicted by the analytical model, artificially scaled to show qualitative agreement.

simulation spectra by tuning the attenuation (δ) and coupling (z) coefficients for both the fundamental TC modes, TE0 (Fig 4b; z = 0.2, $\delta$ = 0.05) and TM0 (Fig 4d; z = 0.4, $\delta$ = 1.0). The NHPM is used to predict patterns in quality factor and spectra deviations for the simulated spectra of both fundamental TE0 (Fig 4b) and TM0 (Fig 4d) TC modes, shown as black dots which closely trace the NHPM with increasing mode number ($M$). These results show that the analytical model successfully captures all physics for the fundamental TM0 and TE0 TC modes. In most cases, deviations from the analytically predicted spectrum can be compensated by introducing frequency dependent design coefficients.

The z-coefficient is directly related to the coupling strength between the ring resonator and waveguide. Large z-coefficients imply larger power exchange between the ring resonator and waveguide leading to weaker confinement of cavity and subsequently a reduced quantity factor outside of the BIC condition, as shown in Fig S2. This can be physically controlled by engineering the geometry of the mirror symmetric cross junction including aspect ratio of the "cross" shaped geometry and radius of curvature at the corners. The z-coefficient is also affected by the spatial distribution of the electric field in the cross junction, which is directly related to the polarization of the waveguide mode. This is responsible for the lower z-coupling coefficient of the TE0 mode (z = 0.2, Fig 4d) compared to the TM0 mode (z = 0.2, Fig 4e), resulting in broader TM0 resonances in the simulated spectra in Fig 4.

`

## Results and Discussions

The Theta Cavity's unique spectral response is determined by the geometric tuning parameter ($\gamma$) which can be engineered according to the physical geometry of the TC. This is demonstrated experimentally with passive Silicon-on-Insulator (SOI) devices at telecom wavelengths (~1550 nm), utilizing a standard single layer process to fabricate fully etched single mode waveguides (500 nm width and 220 nm height) and tapered grating couplers, (more information provided in the supplemental section). A Scanning Electron Microscopy (SEM) image of the circular theta cavity and grating couplers on SOI is shown in Fig 5a. The measured transmission spectrum reveals the presence of sharp and deep mode features in unique spectra patterns, reminiscent of narrow band-drop filter response [52–54]. We observe that TC modes have characteristically large extinction ratios compared to state-of-the-art ring resonator devices, typically ranging from -20 dB to -30 dB, resulting from the inherently efficient power transfer at the interferometric MSC junction [6,7,28]. In this study, the maximum observed quality factors of TC modes in measured spectra typically range from 10,000 – 15,000, limited by the extrinsic factors associated with device design and fabrication [6,7,28]. Although theoretical quality factors for Quasi-BIC modes are predicted to approach infinity, the quality factors of observable TC modes are limited by unavoidable leakage pathways into the far-field in real SOI devices, captured by the attenuation coefficient ($\delta$). For the devices used in this study, non-idealities in the fabrication of the SOI waveguides, such as the lack of a cladding layer and line edge roughness, lead to radiative losses which limit the realizable quality factor for TC qBICs. In addition, the cross junction supports supermode formation, shown in the simulated electric field in Fig 4ac, which creates a radiative pathway into the far-field. The supermode coupling to the far-field doesn't preclude the formation of Theta Cavity BICs since the qBIC to BIC transition is robust to modest attenuation ($\delta > 0$), as shown in Fig 2f, and the far-field radiation pathways are isolated from the continuum of propagating waveguide modes. Radiative loss due to far-field coupling in the Theta Cavity is not a fundamental limitation since it can be significantly suppressed through rigorous engineering of the physical cross junction and the material platform.

Theta Cavity Quasi-BICs exhibit significantly reduced leakage at the junctions by suppressing the backward propagation pathway which allows more light to be transmitted through the waveguide. This results in "shallow" modes with comparatively smaller extinction coefficients and higher quality factor than other modes in the spectra; as shown in the analytical simulations in Fig S3. For larger diameter rings (D = 35 um), the measured transmission spectrum (Fig 5b) exhibits shallow modes with higher quality factors and a distinctive spectral pattern, characteristic of quasi-BICs which manifest under lower attenuation coefficient. In Fig 5cd, the bandwidth and normalized frequency shift extracted from the measured spectra pattern show agreement with spectral trends predicted by the analytical model ($z = 0.2, \delta = 10^{-4}, \gamma = 0.629$). We use these distinct features to identify qBICs in the spectra, validated by the expected qBIC periodicity predicted by the analytical model. For smaller diameter rings (D = 8.5 um), the measured transmission spectra (Fig 5e) and the analytical spectra (Fig 5f), fitted with a higher attenuation coefficient ($\delta = 0.1$), strongly agree and shows unambiguous evidence of a BIC transition. Notice that the occurrence of the BIC transition is strongly emergent in the spectra for the smaller diameter TC, arising from the opening of the NHPM due to its higher attenuation coefficient (refer to Fig 2f). Whereas the larger diameter TC devices have lower attenuation coefficient leading to a narrow transition in the NHPM and are therefore more likely to show as a qBIC in the spectra. These empirical observations together with the analytical band structure provide compelling evidence of the non-Hermitian nature of TC modes, where the non-trivial transitions are protected from attenuation in SOI devices.

The TC showcases remarkably high stability and robustness to non-idealities in monolithic fabrication (imperfect waveguides) and environmental perturbations (temperature dependent

`

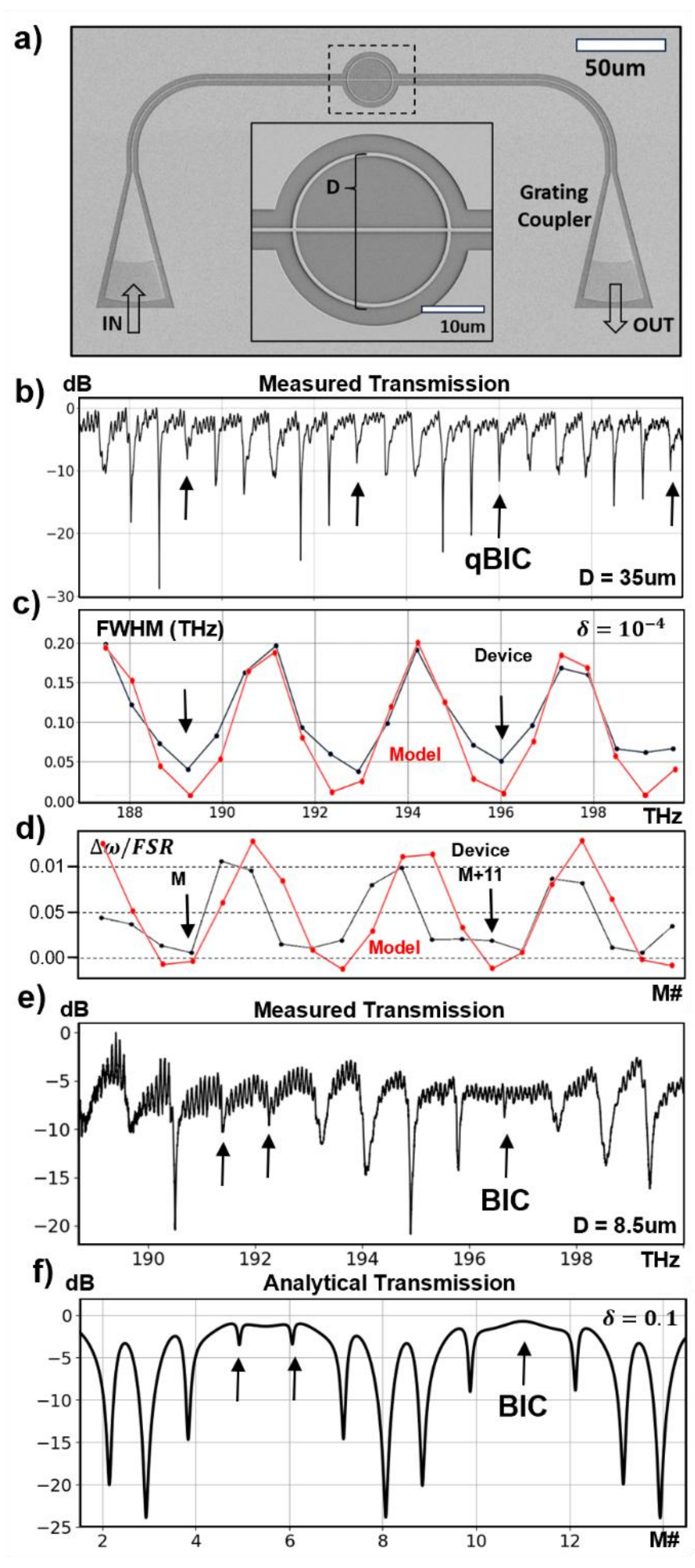

**Fig 5: Circular Theta Cavity Results.** (a) Scanning Electron Microscopy image of a representative Circular Theta Cavity with grating couplers on SOI substrate. (b) The measured extinction spectra for a larger diameter circular Theta Cavity, D = 35 um. (c) The full width half max, FWHM, of modes extracted from the measured spectra and fitted analytical model, $z = 0.2, \delta = 10^{-4}, \gamma = 0.629$; arrows highlight Quasi-BICs in the spectra. (d) The relative spectral shift ($\Delta\omega_M/\Delta\omega_{FSR}$) for modes extracted from measure spectra. (e) The measured extinction spectra for a larger diameter circular Theta Cavity, D = 35 um. (f) The analytical transmission spectra assuming a larger loss coefficient, $z = 0.2, \delta = 0.\ 1, \gamma = 2/\pi$; arrows highlight BICs in the spectra. Artificially flattening of all measured spectra is performed by subtracting the background to account for device variability.

`

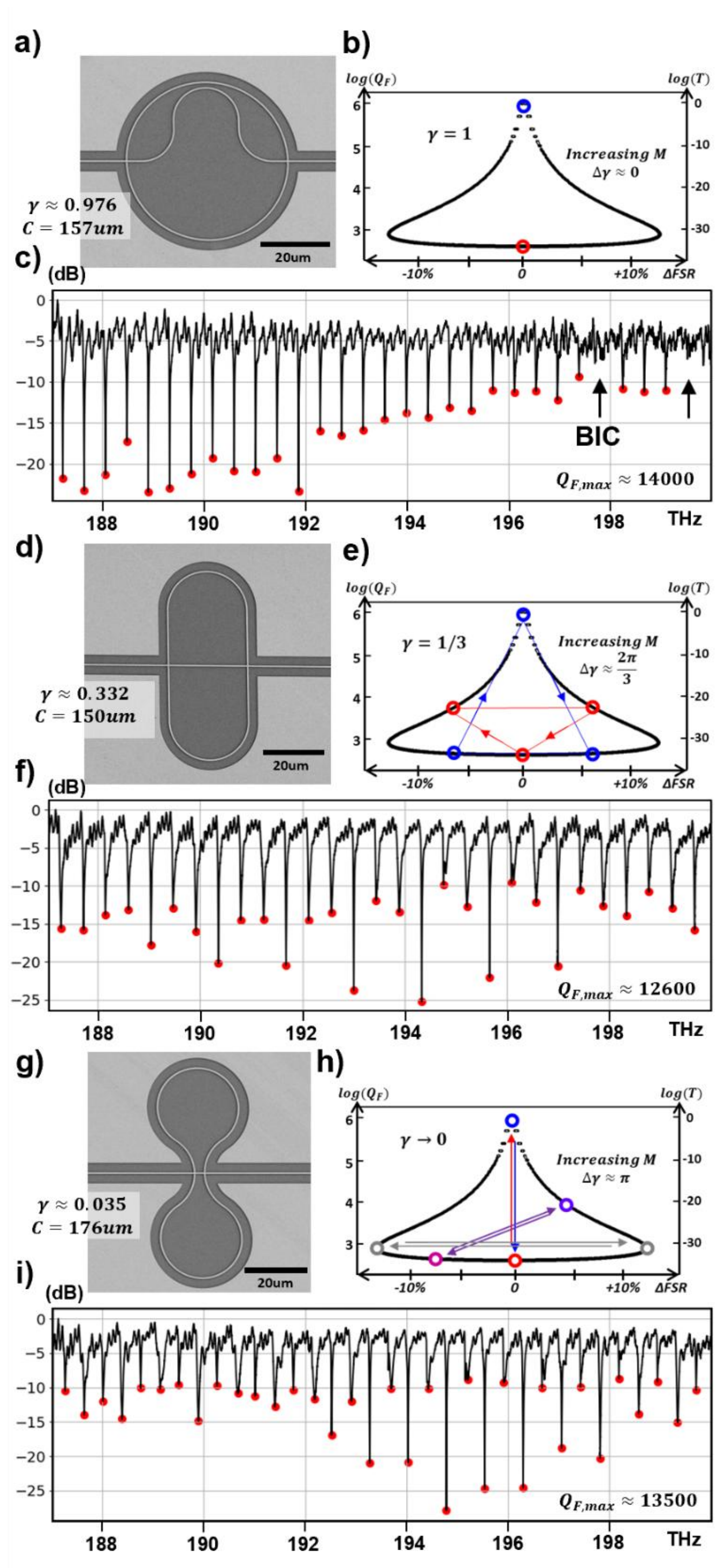

**Fig 6: Engineering Spectra Pattern Behavior.** The experimental results for the balanced (g=1) (a-c), periodic (g=1/3) (d-f), and oscillating (g=0) (g-i) Theta Cavity devices. (a,d,g) A scanning electron microscope of fabricated SOI devices is shown, the design coefficients is shown in inset. (b,e,h) The NHPM illustrates the cyclical behavior of spectral pattern for a given gamma value respectively predicting quality factor, $\log(Q_F)$, and extinction (dB) changes with increasing mode number, M. (c,f,i) The measured extinction spectra plotted in dB scale is shown, the extinction peaks are marked as red points to showcase unique spectra behavior.

`

refractive index drift) due to its symmetric phase condition and the absence of a physical barrier to mediate coupling. Spectra from various circular TCs (Fig S4) display the same spectral patterns consistent with predicted trends despite variability among devices, highlighting robustness to fabrication non-idealities. We also preformed temperature dependent measurements which show temperature stability of the mode. Temperature-dependent transmission spectra of engineered TCs with different gamma values (Fig S5) reveal stable bandwidth and extinction spectral pattern under changing temperature ($\Delta T = 40^{o}$ C), which confirms robustness to environmental perturbations. A sensitivity analysis using a simulation model of the circular TCs was performed to evaluate the spectral response to typical geometric variability, demonstrating notable robustness under fabrication tolerances commonly encountered in modern processing (Fig. S6).

Balanced TC ($\gamma \approx 1$). An SEM image of a balanced path Theta Cavity (Fig 6a) displays a TC with an inner waveguide (path 1) and outer ring resonator (path 2) that have identical effective path lengths. The challenge with this design was maintaining a >2.5 um gap between the parallel waveguides to avoid coupled line effects. When both paths have equal lengths, the phase difference is effectively zero creating a monotonic pattern of sharp resonances in which the mode character changes gradually with increasing mode number, as illustrated by the NHPM (Fig 6b upper). In Fig 6c, the measured spectra reveal a gradually varying spectral pattern of high-Q modes, trending towards decreasing extinction with increasing mode number eventually crosses through the BIC transition. We observed different mode characteristics within the spectral patterns of other balanced TC devices due to slight differences in their geometric parameter, exhibiting a monotonic series of sharp modes and broader modes shown in Fig S7. Both trends show strong agreement with predicted spectral pattern characteristics for the balanced TC.

Cyclical TC ($\gamma \approx 1/3$). When the gamma parameter can be described as a fraction of rational numbers, the spectral pattern will have a distinct cyclical feature as the mode selection moves around the NHPM in a periodic pattern. Here design complexity is related to engineering precise gamma values by considering difference in effective index of curved and straight waveguides which comprise the cavity structure. In Fig 6d, an SEM image of a cyclical Theta Cavity, designed such that $\gamma \approx 1/3$, creates repeating spectral features every third mode, as illustrated by the NHPM (Fig 6d). As predicted, the measured spectral pattern (Fig 6e) shows a cyclical trend in extinction ratio and bandwidth that repeats every third mode. Notice the pattern cycle does drift slowly with increasing mode number due to the slight discrepancy between the target gamma value of the device design and the actual gamma value of the device.

Binary TC ($\gamma \approx 0$). When the gamma parameter approaches zero, resonances alternate between the two extremes on the NHPM corresponding to an effective $\pi$ phase difference between modes in the spectral pattern. However, since the architecture requires both paths to have finite length, $\gamma = 0$ is not physically realizable, which implies the alternating mode behaviors will gradually shift around the NHPM (Fig 6g). This creates a design challenge for engineering gamma near zero, since the Binary TC design requires a very short middle waveguide (path 1), limited by coupled line effects, and a large circumference ring resonator (path 2), limited by propagation loss and target free spectra range. Our design solution, shown as an SEM image in Fig 6g, resembles an "hourglass" shape with an approximate gamma value $\gamma \approx 0.035$. The measured spectral pattern (Fig 6h) shows strong evidence of "binary" spectral pattern showcasing alternating "sharp" and "broad" mode characteristics with increasing mode number, which agree with spectral trends predicted by the analytical model.

`

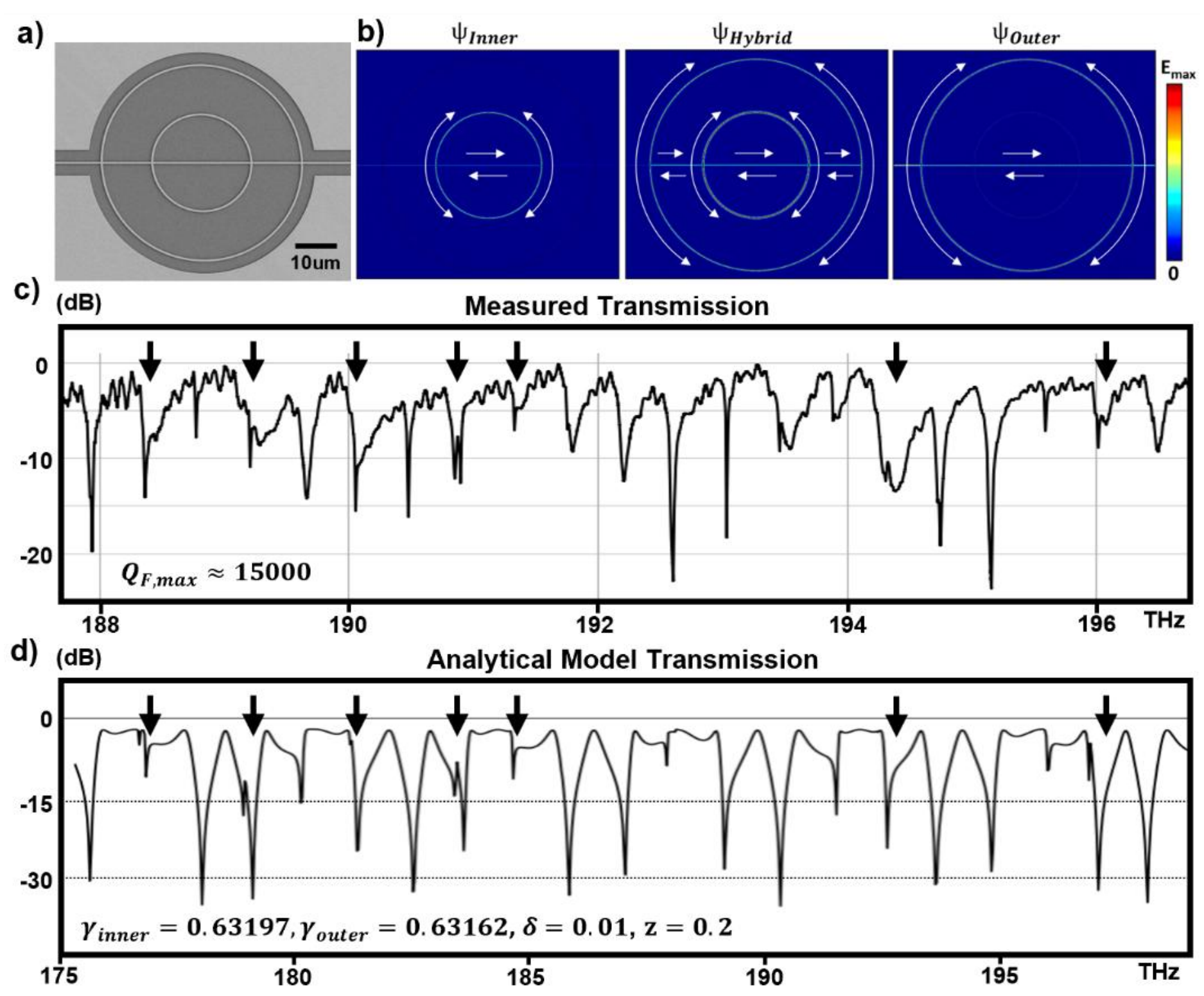

**Fig 7: Nested Theta Cavity Architecture.** (a) A scanning electron microscopy image of the nested-ring Theta Cavity device on SOI, $D_{inner}$= 25 um and $D_{outer}$ = 50 um. (b) The simulated Electric field map reveals mode confinement to the inner ring (left), both rings (middle), and outer ring (right) respectively. (c) The measured extinction spectra is plotted. (d) The extinction spectra predicted by analytical model with tuning parameters chosen to qualitatively show similar mode features and patterns highlighted by black arrows.

## Coupled Resonators

Optical resonator-resonator coupling has become a growing area of interest owing to their ability to support exotic light-matter interactions for a variety of photonic platforms, including Non-Hermitian Optics [49,55,56], Resonator Lattices [50,57–60], synthetic dimensions [59–64], wavelength conversion [65,66], and Quantum Photonics [67–69]. Near-field coupling is a commonly used approach to facilitate resonator-resonator coupling, which requires engineering multiple ring resonators in close physical proximity to each other [52,70–77]. The near-field coupling of multiple high quality factor modes increases sensitivity to perturbation and variability which can lead to system instability and significantly limits scalability [19,52,78]. In the Nested Theta Cavity (NTC) architecture, a smaller ring resonator is placed inside a larger outer ring resonator, the shared middle waveguide path facilitates a correlated phase relationship between the concentric rings. A novel feature of the Nested Theta Cavity is long-range resonator-resonator strong coupling without any limitation in physical proximity typically associated with near-field coupling. The dual nested ring architecture has seven distinct physical paths, each representing a unique parameter set in the multi-dimensional band structure. When both ring resonators satisfy the resonant TC condition near the same frequency, their relative phase relationship becomes strongly correlated leading to band hybridization with unique synthetic dimensions. This cavity design paradigm allows for scaling of strongly correlated physical and synthetic dimensions with the incorporation of many nested rings in a single cavity structure. A distinct advantage of

`

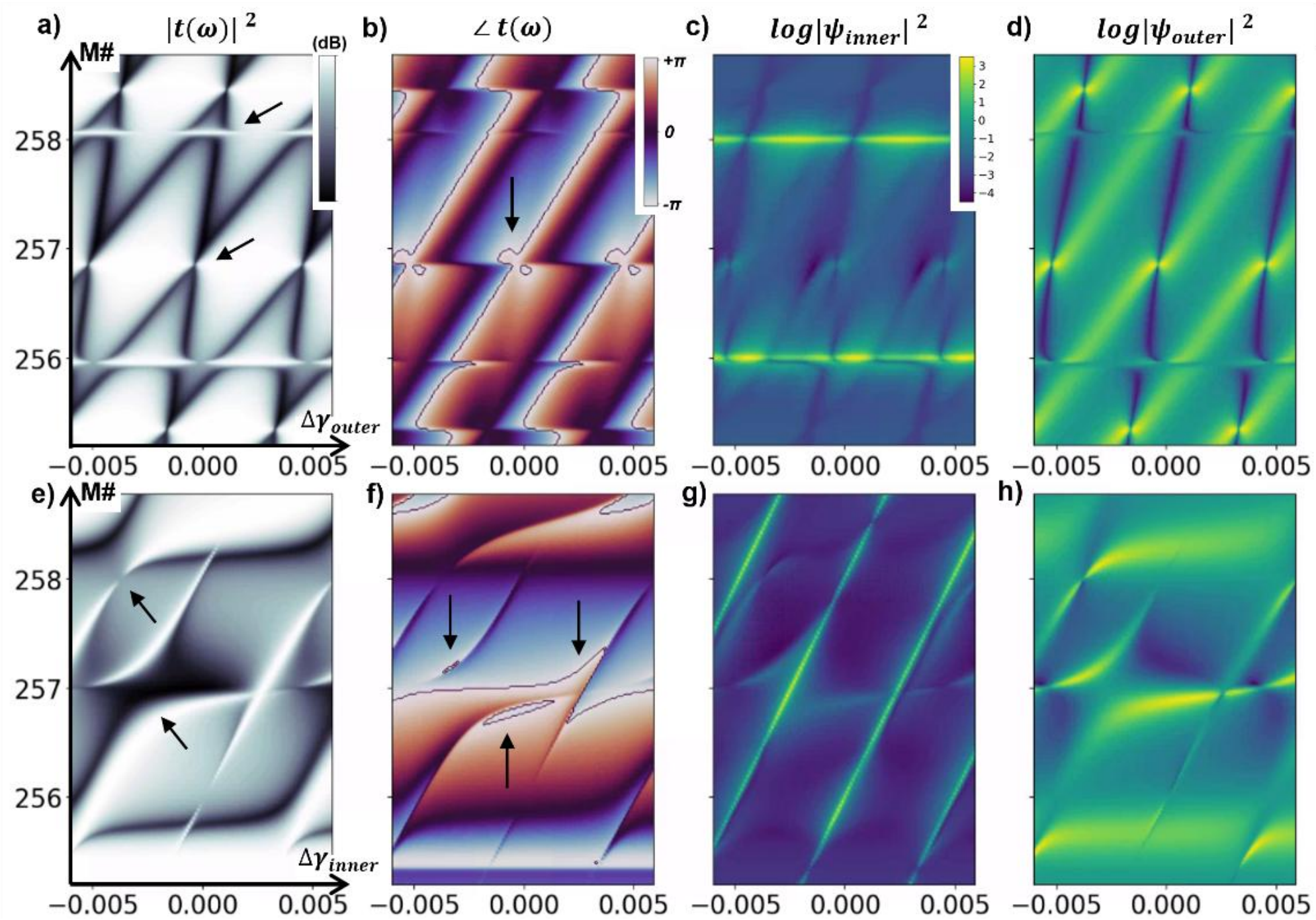

**Fig 8: NTC Band Structure Hybridization.** Dispersion relation predicted by analytical model for the Nested Theta Cavity (NTC) is plotted as a false color contour plot for the dynamic outer ring phase parameter $\Delta\gamma_{outer}$ (a,b,c,d) and dynamic inner ring phase parameter $\Delta\gamma_{inner}$ (e,f,g,h). (a,e) The simulated transmission spectra, (b,f) relative phase relation (arrows highlight both asymptotic non-trivial transitions as well as FW-BICs), (c,g) electric field enhancement in inner ring, (d,h) electric field enhancement in outer ring.

the NTC is the inherent topological robustness and efficient power transfer in the coupled resonator system does not diminish with increasing complexity.

In this investigation we focus on the phase parameters of the inner and outer ring resonator pathways to study resonator-resonator coupling leading to band structure hybridization. Our NTCs are designed with a diameter of the inner ring is half that of the outer ring ($D_{IN}$ = 25 um, $D_{OUT}$ = 50 um), ensuring the strongly overlapping NHPM for every second mode in the spectral pattern. An SEM images of the NTC device used for experimental measurements is shown in Fig 7a, showcasing its unique concentric-ring architecture. The simulation model (Fig 7b) of the normalized electric field for strongly coupled modes reveals three distinctive mode confinement states: *Inner-Ring* mode dominant, *Outer-Ring* mode dominant, and *Hybrid-Ring* modes with strong confinement in both ring resonators.

For the NTC analytical model, the interface-continuity equations are applied to all four MSC junctions in the dual ring architecture to produce closed form equations to describe the wavefunctions in the system. The nested ring analytical model can be expanded for any number of concentric nested rings using a Matrix transfer method, described in more detail in supplemental section. The NTC analytical model used for this study is derived by applying design constraints representing the correlated phase relationship between the two ring resonators linked by the middle path. The spectral trends and features observed in the measured NTC spectra (Fig 7c) agree strongly with the nested ring analytical spectra (Fig. 7d) after qualitatively fitting using design coefficients ($\gamma_{inner} = 0.63197, \gamma_{outer} = 0.63162, \delta = 0.01$, z = 0.2). In both the measured and analytical transmission spectrum, spectral overlap occurs

`

every other resonance appearing as a single peak in the spectra (crossing), as a bimodal peak (anti-crossing), or a weaker mode signature (qBICs).

The optical dispersion relation of the NTC can be mapped by plotting the spectral response (y-axis) against the tuning parameter ($\gamma_{inner}$ and $\gamma_{outer}$) for the individual ring resonators (x-axis), introducing a phase difference between the ring resonator paths. The dispersion relation shown in Fig 8, was mapped numerically from the nested ring analytical model by sweeping the geometric tuning parameter for the inner and outer ring independently. The simulated dispersion relation shown in Fig S8, were obtained by tuning the refractive index of the outer ring while keeping the index of the inner ring constant. Both the analytical and simulated dispersion relations reveal the same band structures and spectral features which further validates our analytical NTC model. This highlights the significant computational advantage of using our closed form equations to predict multidimensional dispersion when compared to numerical electromagnetic simulations, which is essential for scaling the complexity of multiring NTC architectures. We hypothesize that the direct interference between the inner and outer nested TC modes results in a Friedrich-Wintgen BIC condition, which is responsible for the emergent mode splitting near the bandgap in Fig 8a, as well as the Fano to BIC transitions and "Anti-Crossings" observed throughout the dispersion in Fig 8e [45,48].

For the Outer Ring tuning parameter ($\gamma_{outer}$), Fig 8a-d, the horizontal bands in the dispersion plots result from the band structure in the inner ring which is kept constant. The "anti-crossing" band hybridization features a bandgap along the frequency and/or index axis (Fig 8a) with bimodal features along the bandgap edges and suppression of the mode features inside the bandgap, while maintaining strong confinement in the inner ring path (Fig 8c). Near the bandgap features, annihilation of mode confinement in one or both rings' manifests as abrupt changes in the spatial confinement of the mode from $\psi_{inner} \leftrightarrow \psi_{outer}$ (Fig 8cd). We also observe "Dirac-crossing" band hybridization which does not exhibit a bandgap (Fig 8a) and features strong mode confinement to the outer ring (Fig 8d) with mode suppression in the inner ring (Fig 8c). Near the "Dirac-crossing" features, the inner ring exhibits weak mode confinement that coincides with the maximum in the outer ring mode leading to the emergence hybrid mode states ($\psi_{hybrid}$). In Fig S9, the transmission dispersion relation is mapped across a large range of mode numbers (M) while tuning the gamma parameter, revealing the correlation between size of the bandgap and the relative phase relation between the rings. The band structure evolution from "Dirac-Crossing" features to an "Anti-Crossing" signature with characteristic bandgap features with changing mode number is emblematic of non-trivial topological bands and FW-BIC formation [44,45,48,79].

For the inner ring tuning parameter ($\gamma_{inner}$), Fig 8f-i, the horizontal bands in the dispersion plots result from the band structure in the outer ring which is kept constant. Our results reveal interesting dynamics including abrupt inversions in propagation direction of the outcoupled mode highlighted by narrowband Fano Resonance line shapes in the transmission dispersion (Fig 8f). The mode confinement maps (Fig 8gh) predict a gradual shift in spatial confinement between the inner and outer rings along the "anti-crossing" bands, with overlapping bands leading to the formation of strongly confined hybrid modes. In the NTC, the selection rules for inner ring modes are independent of outer ring coupling, whereas the outer ring's resonant condition is strongly influenced by coupling with the inner ring. The inequivalence between the dynamic inner ring and dynamic outer ring arises from the difference in their respective resonant condition which depends on the phase relationship between the middle waveguide paths and ring resonator. Notice that tuning the inner ring index creates nontrivial changes to the phase of the paths connecting the inner and outer ring resonators, subsequentially modifying the resonant condition of the outer ring. This is responsible for the observed emergent splitting of asymptotic bands in the phase dispersion marked by black arrows in Fig 8g. In contrast, since the outer ring externally coupled to the inner TC, tuning to the outer ring index creates linear

`

shifts in the phase condition of the inner ring resulting in trivial effects on the NTC resonant condition and periodic features in the dispersion plots (Fig 8a-d).

## Conclusion

We have introduced the Theta Cavity as a new interferometric resonator architecture which expands the design parameter space for integrated photonic beyond the constraints imposed by near-field coupling. By leveraging mirror-symmetric interferometric cross junctions and phase-engineered multipath interference, the Theta Cavity exhibits efficient power transfer supporting high extinction modes that are intrinsically robust to fabrication variability and environmental perturbations. The observation of parametric BIC modes and their unifying description through a Non-Hermitian Phase Manifold reveal previously inaccessible cavity dynamics and provide new physical insight into phase-driven resonator systems.

The analytical framework developed here offers a predictive and computationally efficient tool for engineering spectral response, quality factor, and band structure; it is shown to accurately capture both single and nested TC behavior. The Theta Cavity architecture provides a new and versatile foundation for exploring topological and non-Hermitian photonics, synthetic dimensions, and strongly coupled resonator lattices, with direct implications for filtering, sensing, nonlinear optics, and quantum photonic platforms.

The Nested Theta Cavity offers a novel design space for the investigation of non-trivial topology in coupled resonators systems**.** The demonstration of phase-mediated long-range strong coupling in Nested Theta Cavity systems establishes a scalable route toward complex resonator networks within a compact and fabrication-tolerant architecture. Our results demonstrate the ability to construct NTC rings and engineer a multidimensional parameter space showing topological nontriviality and band hybridization in the simplest coupled case for two co-resonant nested rings. These results are strong indicators that more complex multidimensional structures are possible, providing us with potential access to new exotic topological behavior. Notice that the nested ring analytical model can be expanded for any number of concentric nested rings using a matrix transfer method, described in supplemental section. This provides a framework to engineer strongly coupled parameter spaces across arrays of ring resonators supporting synthetic dimensions, higher dimensional topological band structure, and exotic non-Hermitian physics.

## Acknowledgments

We appreciate Prof. William L Wilson at Harvard University for valuable discussion. We are grateful for the financial support provided by U.S. Army Research. Additionally, we acknowledge the IBM HBCU Quantum Center for supporting this work.
U.S. Army Research Office (W911NF-24-1-0217)

## Data availability

The data that support the plots within this paper and other findings of this study are available from the corresponding author upon reasonable request.

`

## References

(1) Xie, Y.; Zhang, M.; Dai, D. Design Rule of Mach-Zehnder Interferometer Sensors for Ultra-High Sensitivity. *Sensors (Switzerland)* **2020**, *20* (9). https://doi.org/10.3390/s20092640.

(2) Yurtsever, G.; Považay, B.; Alex, A.; Zabihian, B.; Drexler, W.; Baets, R. Photonic Integrated Mach-Zehnder Interferometer with an on-Chip Reference Arm for Optical Coherence Tomography. *Biomed. Opt. Express* **2014**, *5* (4). https://doi.org/10.1364/boe.5.001050.

(3) Hiraki, T.; Fukuda, H.; Yamada, K.; Yamamoto, T. Small Sensitivity to Temperature Variations of Si-Photonic Mach–Zehnder Interferometer Using Si and SiN Waveguides. *Front. Mater.* **2015**, *2*. https://doi.org/10.3389/fmats.2015.00026.

(4) Lin, X.; Rivenson, Y.; Yardimci, N. T.; Veli, M.; Luo, Y.; Jarrahi, M.; Ozcan, A. All-Optical Machine Learning Using Diffractive Deep Neural Networks. *Science (1979).* **2018**, *361* (6406). https://doi.org/10.1126/science.aat8084.

(5) Ramelow, S.; Farsi, A.; Vernon, Z.; Clemmen, S.; Ji, X.; Sipe, J. E.; Liscidini, M.; Lipson, M.; Gaeta, A. L. Strong Nonlinear Coupling in a Si3 N4 Ring Resonator. *Phys. Rev. Lett.* **2019**, *122* (15). https://doi.org/10.1103/PhysRevLett.122.153906.

(6) Prinzen, A.; Waldow, M.; Kurz, H. Fabrication Tolerances of SOI Based Directional Couplers and Ring Resonators. *Opt. Express* **2013**, *21* (14). https://doi.org/10.1364/oe.21.017212.

(7) Baehr-Jones, T.; Hochberg, M.; Walker, C.; Scherer, A. High-Q Ring Resonators in Thin Silicon-on-Insulator. *Appl. Phys. Lett.* **2004**, *85* (16), 3346–3347. https://doi.org/10.1063/1.1781355.

(8) Chremmos, I.; Uzunoglu, N. Reflective Properties of Double-Ring Resonator System Coupled to a Waveguide. *IEEE Photonics Technology Letters* **2005**, *17* (10), 2110–2112. https://doi.org/10.1109/LPT.2005.854346.

(9) Wang, X.; Wang, J.; Yao, Y.; Xiao, S.; Song, Q.; Xu, K. Efficient and High-Speed Coupling Modulation of Silicon Racetrack Ring Resonators at 2 Mm Waveband. *Opt. Lett.* **2024**, *49* (8). https://doi.org/10.1364/ol.518729.

(10) Bogaerts, W.; de Heyn, P.; van Vaerenbergh, T.; de Vos, K.; Kumar Selvaraja, S.; Claes, T.; Dumon, P.; Bienstman, P.; van Thourhout, D.; Baets, R. Silicon Microring Resonators. *Laser and Photonics Reviews*. January 2012, pp 47–73. https://doi.org/10.1002/lpor.201100017.

(11) Xu, H.; Qin, Y.; Hu, G.; Tsang, H. K. Million-Q Integrated Fabry-Perot Cavity Using Ultralow-Loss Multimode Retroreflectors. *Photonics Res.* **2022**, *10* (11). https://doi.org/10.1364/prj.470644.

(12) Wu, J.; Moein, T.; Xu, X.; Ren, G.; Mitchell, A.; Moss, D. J. Micro-Ring Resonator Quality Factor Enhancement via an Integrated Fabry-Perot Cavity. *APL Photonics* **2017**, *2* (5). https://doi.org/10.1063/1.4981392.

(13) Dyakov, S. A.; Salakhova, N. S.; Ignatov, A. V.; Fradkin, I. M.; Panov, V. P.; Song, J. K.; Gippius, N. A. Chiral Light in Twisted Fabry–Pérot Cavities. *Adv. Opt. Mater.* **2024**, *12* (12). https://doi.org/10.1002/adom.202302502.

(14) Cheng, H.; Xiang, C.; Jin, N.; Kudelin, I.; Guo, J.; Heyrich, M.; Liu, Y.; Peters, J.; Ji, Q. X.; Zhou, Y.; Vahala, K. J.; Quinlan, F.; Diddams, S. A.; Bowers, J. E.; Rakich, P. T. Harnessing Micro-Fabry–Pérot Reference Cavities in Photonic Integrated Circuits. *Nat. Photonics* **2025**, *19* (9). https://doi.org/10.1038/s41566-025-01701-5.

(15) Atvars, A. Analytical Description of Resonances in Fabry–Perot and Whispering Gallery Mode Resonators. *Journal of the Optical Society of America B* **2021**, *38* (10), 3116. https://doi.org/10.1364/josab.419993.

(16) Xu, D.-X.; Densmore, A.; Waldron, P.; Lapointe, J.; Post, E.; Delâge, A.; Janz, S.; Cheben, P.; Schmid, J. H.; Lamontagne, B. High Bandwidth SOI Photonic Wire Ring Resonators Using MMI Couplers. *Opt. Express* **2007**, *15* (6). https://doi.org/10.1364/oe.15.003149.

(17) Kharel, P.; Chu, Y.; Mason, D.; Kittlaus, E. A.; Otterstrom, N. T.; Gertler, S.; Rakich, P. T. Multimode Strong Coupling in Cavity Optomechanics. *Phys. Rev. Appl.* **2022**, *18* (2). https://doi.org/10.1103/PhysRevApplied.18.024054.

(18) Huang, W.-P. Coupled-Mode Theory for Optical Waveguides: An Overview. *Journal of the Optical Society of America A* **1994**, *11* (3). https://doi.org/10.1364/josaa.11.000963.

(19) Tseng, C.-W.; Tsai, C.-W.; Lin, K.-C.; Lee, M.-C.; Chen, Y.-J. Study of Coupling Loss on Strongly-Coupled, Ultra Compact Microring Resonators. *Opt. Express* **2013**, *21* (6). https://doi.org/10.1364/oe.21.007250.

(20) Chang, T. H.; Zhou, X.; Zhu, M.; Fields, B. M.; Hung, C. L. Efficiently Coupled Microring Circuit for On-Chip Cavity QED with Trapped Atoms. *Appl. Phys. Lett.* **2020**, *117* (17). https://doi.org/10.1063/5.0023464.

(21) Hinney, J.; Kim, S.; Flatt, G. J. K.; Datta, I.; Alù, A.; Lipson, M. Efficient Excitation and Control of Integrated Photonic Circuits with Virtual Critical Coupling. *Nat. Commun.* **2024**, *15* (1). https://doi.org/10.1038/s41467-024-46908-2.

(22) Jannesari, R.; Pühringer, G.; Stocker, G.; Grille, T.; Jakoby, B. Design of a High Q-Factor Label-Free Optical Biosensor Based on a Photonic Crystal Coupled Cavity Waveguide. *Sensors* **2024**, *24* (1). https://doi.org/10.3390/s24010193.

(23) van Niekerk, M.; Cheng, C.; Leake, G.; Coleman, D.; Fanto, M. L.; Preble, S. F. High Extinction Ratio Microring Modulator. In *Optics InfoBase Conference Papers*; 2022.

`

(24) van Niekerk, M.; Cheng, C.; Leake, G.; Coleman, D.; Fanto, M. L.; Preble, S. F. High Extinction Ratio Microring Modulator. In *2022 Conference on Lasers and Electro-Optics, CLEO 2022 - Proceedings*; 2022. https://doi.org/10.1364/cleo_si.2022.sth4k.5.

(25) Kar, S.; Singhal, S.; Paithankar, P.; Varshney, S. K. Microring Resonators and Its Applications. *Indian Journal of Pure and Applied Physics* **2023**, *61* (7). https://doi.org/10.56042/ijpap.v61i7.110.

(26) Prinzen, A.; Bolten, J.; Waldow, M.; Kurz, H. Study on Fabrication Tolerances of SOI Based Directional Couplers and Ring Resonators. *Microelectron. Eng.* **2014**, *121*. https://doi.org/10.1016/j.mee.2014.03.019.

(27) Kustura, K.; Gonzalez-Ballestero, C.; Sommer, A. D. L. R.; Meyer, N.; Quidant, R.; Romero-Isart, O. Mechanical Squeezing via Unstable Dynamics in a Microcavity. *Phys. Rev. Lett.* **2022**, *128* (14). https://doi.org/10.1103/PhysRevLett.128.143601.

(28) Niehusmann, J.; Vörckel, A.; Bolivar, P. H.; Wahlbrink, T.; Henschel, W.; Kurz, H. Ultrahigh-Quality-Factor Silicon-on-Insulator Microring Resonator. *Opt. Lett.* **2004**, *29* (24), 2861. https://doi.org/10.1364/ol.29.002861.

(29) Elshaari, A. W.; Aboketaf, A.; Preble, S. F. Controlled Storage of Light in Silicon Cavities. *Opt. Express* **2010**, *18* (3). https://doi.org/10.1364/oe.18.003014.

(30) Preble, S.; Bergman, B.; Carpenter, L. G.; Chrostowski, L.; Dikshit, A.; Fanto, M.; Lin, W.; van Niekerk, M.; Uddin, M. R.; Sundaram, V. S. S. Passive Silicon Photonic Devices. In *Integrated Photonics for Data Communication Applications*; 2023. https://doi.org/10.1016/B978-0-323-91224-2.00001-1.

(31) Kim, C. S.; Satanin, A. M. Coherent Resonant Transmission in Temporally Periodically Driven Potential Wells: The Fano Mirror. *Journal of Physics Condensed Matter* **1998**, *10* (47). https://doi.org/10.1088/0953-8984/10/47/010.

(32) Lan, Z.; Chen, M. L. N.; Gao, F.; Zhang, S.; Sha, W. E. I. A Brief Review of Topological Photonics in One, Two, and Three Dimensions. *Reviews in Physics*. 2022. https://doi.org/10.1016/j.revip.2022.100076.

(33) Yuan, L.; Lu, Y. Y. Bound States in the Continuum on Periodic Structures: Perturbation Theory and Robustness. *Opt. Lett.* **2017**, *42* (21). https://doi.org/10.1364/ol.42.004490.

(34) Amrani, M.; Khattou, S.; Al-Wahsh, H.; Rezzouk, Y.; Boudouti, E. H. El; Ghouila-Houri, C.; Talbi, A.; Akjouj, A.; Dobrzynski, L.; Djafari-Rouhani, B. Bound States in the Continuum and Fano Resonances in Photonic and Plasmonic Loop Structures. *Opt. Quantum Electron.* **2022**, *54* (9). https://doi.org/10.1007/s11082-022-03991-3.

(35) Abdel-Ghaffar, E. A.; Dobrzyński, L.; Al-Wahsh, H.; Akjouj, A. Bound States in the Continuum and Long-Lived Electronic Resonances in Two-Tangent Loops Cavity. *Sci. Rep.* **2025**, *15* (1). https://doi.org/10.1038/s41598-025-26444-9.

(36) Gao, Y.; Ge, J.; Sun, S.; Shen, X. Dark Modes Governed by Translational-Symmetry-Protected Bound States in the Continuum in Symmetric Dimer Lattices. *Results Phys.* **2022**, *43*. https://doi.org/10.1016/j.rinp.2022.106078.

(37) Bogdanov, A. A.; Koshelev, K. L.; Kapitanova, P. V.; Rybin, M. V.; Gladyshev, S. A.; Sadrieva, Z. F.; Samusev, K. B.; Kivshar, Y. S.; Limonov, M. F. Bound States in the Continuum and Fano Resonances in the Strong Mode Coupling Regime. *Advanced Photonics* **2019**, *1* (1). https://doi.org/10.1117/1.AP.1.1.016001.

(38) Hsu, C. W.; Zhen, B.; Stone, A. D.; Joannopoulos, J. D.; Soljacic, M. Bound States in the Continuum. *Nature Reviews Materials*. Nature Publishing Group July 19, 2016. https://doi.org/10.1038/natrevmats.2016.48.

(39) Kang, M.; Liu, T.; Chan, C. T.; Xiao, M. Applications of Bound States in the Continuum in Photonics. *Nature Reviews Physics* **2023**, 659–678. https://doi.org/10.1038/s42254-023-00642-8.

(40) Koshelev, K. L.; Sadrieva, Z. F.; Shcherbakov, A. A.; Kivshar, Yu. S.; Bogdanov, A. A. Bound States in the Continuum in Photonic Structures. *Physics-Uspekhi* **2023**, *66* (05). https://doi.org/10.3367/ufne.2021.12.039120.

(41) Qin, H.; Shi, X.; Ou, H. Exceptional Points at Bound States in the Continuum in Photonic Integrated Circuits. *Nanophotonics* **2022**, *11* (21). https://doi.org/10.1515/nanoph-2022-0420.

(42) Kim, M.; Jacob, Z.; Rho, J. Recent Advances in 2D, 3D and Higher-Order Topological Photonics. *Light: Science and Applications*. 2020. https://doi.org/10.1038/s41377-020-0331-y.

(43) Ding, K.; Fang, C.; Ma, G. Non-Hermitian Topology and Exceptional-Point Geometries. *Nature Reviews Physics*. 2022. https://doi.org/10.1038/s42254-022-00516-5.

(44) Huang, L.; Xu, L.; Powell, D. A.; Padilla, W. J.; Miroshnichenko, A. E. Resonant Leaky Modes in All-Dielectric Metasystems: Fundamentals and Applications. *Physics Reports*. 2023. https://doi.org/10.1016/j.physrep.2023.01.001.

(45) Wang, J.; Li, P.; Zhao, X.; Qian, Z.; Wang, X.; Wang, F.; Zhou, X.; Han, D.; Peng, C.; Shi, L.; Zi, J. Optical Bound States in the Continuum in Periodic Structures: Mechanisms, Effects, and Applications. *Photonics Insights*. 2024. https://doi.org/10.3788/PI.2024.R01.

(46) Koshelev, K.; Lepeshov, S.; Liu, M.; Bogdanov, A.; Kivshar, Y. Asymmetric Metasurfaces with High- Q Resonances Governed by Bound States in the Continuum. *Phys. Rev. Lett.* **2018**, *121* (19). https://doi.org/10.1103/PhysRevLett.121.193903.

(47) Cong, L.; Singh, R. Symmetry-Protected Dual Bound States in the Continuum in Metamaterials. *Adv. Opt. Mater.* **2019**, *7* (13). https://doi.org/10.1002/adom.201900383.

`

(48) Huang, L.; Xu, L.; Rahmani, M.; Neshev, D.; Miroshnichenko, A. E. Pushing the Limit of High-Q Mode of a Single Dielectric Nanocavity. *Advanced Photonics* **2021**, *3* (1). https://doi.org/10.1117/1.AP.3.1.016004.

(49) Parto, M.; Liu, Y. G. N.; Bahari, B.; Khajavikhan, M.; Christodoulides, D. N. Non-Hermitian and Topological Photonics: Optics at an Exceptional Point. *Nanophotonics* **2020**, *10* (1). https://doi.org/10.1515/nanoph-2020-0434.

(50) Ao, Y.; Hu, X.; You, Y.; Lu, C.; Fu, Y.; Wang, X.; Gong, Q. Topological Phase Transition in the Non-Hermitian Coupled Resonator Array. *Phys. Rev. Lett.* **2020**, *125* (1). https://doi.org/10.1103/PhysRevLett.125.013902.

(51) Friedrich, H.; Wintgen, D. Interfering Resonances and Bound States in the Continuum. *Phys. Rev. A (Coll. Park).* **1985**, *32* (6). https://doi.org/10.1103/PhysRevA.32.3231.

(52) Barwicz, T.; Popovic, M. A.; Rakich, P. T.; Watts, M. R.; Haus, H. A.; Ippen, E. P.; Smith, H. I. Microring-Resonator-Based Add-Drop Filters in SiN: Fabrication and Analysis. *Opt. Express* **2004**, *12* (7). https://doi.org/10.1364/opex.12.001437.

(53) Zhang, A.; Yang, X.; Wang, J. Design of Channel Drop Filters Based on Photonic Crystal with a Dielectric Column with Large Radius inside Ring Resonator. *Photonics* **2024**, *11* (6). https://doi.org/10.3390/photonics11060554.

(54) Morichetti, F.; Milanizadeh, M.; Petrini, M.; Zanetto, F.; Ferrari, G.; de Aguiar, D. O.; Guglielmi, E.; Sampietro, M.; Melloni, A. Polarization-Transparent Silicon Photonic Add-Drop Multiplexer with Wideband Hitless Tuneability. *Nat. Commun.* **2021**, *12* (1). https://doi.org/10.1038/s41467-021-24640-5.

(55) Li, A.; Wei, H.; Cotrufo, M.; Chen, W.; Mann, S.; Ni, X.; Xu, B.; Chen, J.; Wang, J.; Fan, S.; Qiu, C. W.; Alù, A.; Chen, L. Exceptional Points and Non-Hermitian Photonics at the Nanoscale. *Nature Nanotechnology*. 2023. https://doi.org/10.1038/s41565-023-01408-0.

(56) Miri, M. A.; Alù, A. Exceptional Points in Optics and Photonics. *Science*. American Association for the Advancement of Science January 4, 2019. https://doi.org/10.1126/science.aar7709.

(57) Leykam, D.; Mittal, S.; Hafezi, M.; Chong, Y. D. Reconfigurable Topological Phases in Next-Nearest-Neighbor Coupled Resonator Lattices. *Phys. Rev. Lett.* **2018**, *121* (2). https://doi.org/10.1103/PhysRevLett.121.023901.

(58) Xu, H. S.; Jin, L. Coherent Resonant Transmission. *Phys. Rev. Res.* **2022**, *4* (3). https://doi.org/10.1103/PhysRevResearch.4.L032015.

(59) Kim, S.; Mann, S. A.; Ni, X.; Alù, A. Slow Light by Topological Routing in Synthetic Dimensions. *ACS Photonics* **2026**, *13* (1). https://doi.org/10.1021/acsphotonics.5c02204.

(60) Lin, Q.; Sun, X. Q.; Xiao, M.; Zhang, S. C.; Fan, S. A Three-Dimensional Photonic Topological Insulator Using a Two-Dimensional Ring Resonator Lattice with a Synthetic Frequency Dimension. *Sci. Adv.* **2018**, *4* (10). https://doi.org/10.1126/sciadv.aat2774.

(61) Lustig, E.; Segev, M. Topological Photonics in Synthetic Dimensions. *Adv. Opt. Photonics* **2021**, *13* (2). https://doi.org/10.1364/aop.418074.

(62) Liu, H.; Yan, Z.; Xiao, M.; Zhu, S. Recent Progress in Synthetic Dimension in Topological Photonics. *Guangxue Xuebao/Acta Optica Sinica*. 2021. https://doi.org/10.3788/AOS202141.0123002.

(63) Ozawa, T.; Price, H. M.; Amo, A.; Goldman, N.; Hafezi, M.; Lu, L.; Rechtsman, M. C.; Schuster, D.; Simon, J.; Zilberberg, O.; Carusotto, I. Topological Photonics. *Rev. Mod. Phys.* **2019**, *91* (1). https://doi.org/10.1103/RevModPhys.91.015006.

(64) Dutt, A.; Lin, Q.; Yuan, L.; Minkov, M.; Xiao, M.; Fan, S. A Single Photonic Cavity with Two Independent Physical Synthetic Dimensions. *Science (1979).* **2020**, *367* (6473). https://doi.org/10.1126/science.aaz3071.

(65) Elshaari, A. W.; Preble, S. F. Engineered Transitions in Photonic Cavities. *Optics and Photonics Journal* **2012**, *02* (04). https://doi.org/10.4236/opj.2012.24030.

(66) Huang, L.; Zhang, W.; Zong, Y.; He, L.; Zhang, X. Hyperbolic Topological Frequency Combs. *ACS Photonics* **2025**, *12* (1). https://doi.org/10.1021/acsphotonics.4c01546.

(67) Yan, Q.; Hu, X.; Fu, Y.; Lu, C.; Fan, C.; Liu, Q.; Feng, X.; Sun, Q.; Gong, Q. Quantum Topological Photonics. *Advanced Optical Materials*. 2021. https://doi.org/10.1002/adom.202001739.

(68) Masuki, K.; Ashida, Y. Berry Phase and Topology in Ultrastrongly Coupled Quantum Light-Matter Systems. *Phys. Rev. B* **2023**, *107* (19). https://doi.org/10.1103/PhysRevB.107.195104.

(69) Downing, C. A.; Toghill, A. J. Quantum Topology in the Ultrastrong Coupling Regime. *Sci. Rep.* **2022**, *12* (1). https://doi.org/10.1038/s41598-022-15735-0.

(70) Chen, H. Z.; Ho, K. L.; Wang, P. H. High-Extinction Photonic Filters by Cascaded Mach–Zehnder Interferometer-Coupled Resonators. *Photonics* **2024**, *11* (11). https://doi.org/10.3390/photonics11111055.

(71) Yan, H.; Xie, Y.; Zhang, L.; Dai, D. Wideband-Tunable on-Chip Microwave Photonic Filter with Ultrahigh-Q U-Bend-Mach–Zehnder-Interferometer-Coupled Microring Resonators. *Laser Photon. Rev.* **2023**, *17* (11). https://doi.org/10.1002/lpor.202300347.

(72) Wu, T.; Lalanne, P. Dissipative Coupling in Photonic and Plasmonic Resonators. *Advanced Photonics* **2025**, *7* (5). https://doi.org/10.1117/1.AP.7.5.056011.

(73) Chang, C.; Sun, Y.; Li, T.; Weng, B.; Zou, Y. Coupling-Controlled Photonic Topological Ring Array. *ACS Photonics* **2024**, *11* (12). https://doi.org/10.1021/acsphotonics.4c01502.

(74) Hotte-Kilburn, A.; Bianucci, P. Integrated Topological Photonics in One Dimension. *Advances in Physics: X*. 2025. https://doi.org/10.1080/23746149.2025.2476417.

`

(75) Selim, M. A.; Anwar, M. Enhanced Q-Factor and Effective Length Silicon Photonics Filter Utilizing Nested Ring Resonators. *Journal of Optics (United Kingdom)* **2023**, *25* (11). https://doi.org/10.1088/2040-8986/acf5fd.

(76) Leykam, D.; Yuan, L. Topological Phases in Ring Resonators: Recent Progress and Future Prospects. *Nanophotonics*. 2020. https://doi.org/10.1515/nanoph-2020-0415.

(77) Wang, X.; Bongiovanni, D.; Wang, Z.; Abdrabou, A.; Hu, Z.; Jukic, D.; Song, D.; Morandotti, R.; El-Ganainy, R.; Chen, Z.; Buljan, H. Construction of Topological Bound States in the Continuum Via Subsymmetry. *ACS Photonics* **2024**, *11* (8). https://doi.org/10.1021/acsphotonics.4c00600.

(78) Butt, M. A. Beyond the Detection Limit: A Review of High-Q Optical Ring Resonator Sensors. *Materials Today Physics*. 2025. https://doi.org/10.1016/j.mtphys.2025.101873.

(79) Azzam, S. I.; Kildishev, A. V. Photonic Bound States in the Continuum: From Basics to Applications. *Advanced Optical Materials*. 2021. https://doi.org/10.1002/adom.202001469.

`

# Supplemental Document

# Photonic Theta Cavity: Engineering Bound States in the Continuum in Topological Resonators Beyond the Limitations of Near-Field Coupling

## 1. Analytical Model Derivations

The Theta Cavity (TC) employs interferometric principles, like a Mach-Zender Interferometer, to facilitate coupling between the waveguide transmission line and a ring resonator cavity. When light interacts within the Mirror Symmetric Cross (MSC) junctions, the energy propagating in the waveguide (path 1) is transmitted or reflected into the ports or couples to the orthogonal ring resonator (path 2). This creates a set of equations to describe the continuity conditions at the interfaces between ring resonator and waveguide paths within the MSC junctions. We use these interface-continuity equations to derive the analytical wave equations for the Theta Cavity and the Nested Theta Cavity designs, shown in detail in this section.

### 1.1 Theta Cavity Analytical Model

The "Normalized Magnitude Poynting vector" refers to a factor that incorporates changes to effective phase velocity and coupling into the paths. The primed Model Assumptions for derivation:

- $z_1$: Normalized Magnitude Poynting vector in path 1.
- $z_2$: Normalized Magnitude Poynting vector in path 2.
- Subscripts $a$ and $b$: forward and backwards-propagating modes
- $\phi_{1a}, \phi_{1b}, \phi_{2a}, \phi_{2b}$ phase difference (absolute) of modes on both ends of cavity.

At the interface, the "continuity equations" result in the following form of equation

$$1 + r = \Psi_{1a} + \Psi_{1b}$$

$$1 + r = \alpha_2(\Psi_{2a} + \Psi_{2b})$$

$$1 + r = \alpha_3(\Psi_{3a} + \Psi_{3b})$$

Where $\alpha_2$ and $\alpha_3$ are parameters specific to the nature of the waveguide modes and relative dimensions of the interface. They can be thought of as the "coupling" constants. The conservation of energy can be rewritten as:

$$n_0(1 - r)(1 + r)^* = n_1(\Psi_{1a} - \Psi_{1b})(\Psi_{1a} + \Psi_{1b})^* + n_2(\Psi_{2a} - \Psi_{2b})(\Psi_{2a} + \Psi_{2b})^* + n_3(\Psi_{3a} - \Psi_{3b})(\Psi_{3a} + \Psi_{3b})^*$$

Exploiting the symmetry of the upper and lower paths (paths 2 and 3), $\Psi_{2a} = \Psi_{3a}$, $\Psi_{2b} = \Psi_{3b}$, $\alpha_2 = \alpha_3 := \alpha$ , and $n_2 = n_3$:

$$n_0(1 - r)(1 + r)^* = n_1(\Psi_{1a} - \Psi_{1b})(\Psi_{1a} + \Psi_{1b})^* + 2n_2(\Psi_{2a} - \Psi_{2b})(\Psi_{2a} + \Psi_{2b})^*$$

Using the interface equations, we can rewrite this as

$$n_0(1 - r)(1 + r)^* = n_1(\Psi_{1a} - \Psi_{1b})(1 + r)^* + 2\frac{n_2(\Psi_{2a} - \Psi_{2b})(1 + r)^*}{\alpha^*}$$

`

$$1 - r = \frac{n_1}{n_0}(\Psi_{1a} - \Psi_{1b}) + 2\frac{n_2}{n_0\alpha^*}(\Psi_{2a} - \Psi_{2b})$$

For the simplicity of the model, it would be convenient to have minimal parameters. We therefore replace the $\Psi_2$ with $\Psi_2' = \alpha_2\Psi_2$ the interface equation involving $\Psi_2$becomes

$$1 + r = \Psi_{2a}' + \Psi_{2b}'$$

While conservation of energy equation becomes

$$1 - r = \frac{n_1}{n_0}(\Psi_{1a} - \Psi_{1b}) + 2\frac{n_2}{n_0\alpha^*\alpha}(\Psi_{2a}' - \Psi_{2b}')$$

$$1 - r = z_1(\Psi_{1a} - \Psi_{1b}) + 2z_2(\Psi_{2a}' - \Psi_{2b}')$$

Where $z_1 = \frac{n_1}{n_0}$ and $z_2 = \frac{n_2}{n_0|\alpha|^2}$. **NOTE:** our model also neglects any effects arising from the finite length of the interface, effectively letting all the intersections occur simultaneously and over infinitesimal length.

Combining our equations, we find the equations and re-arranging so that there are 3 on each side:

$$\begin{bmatrix} 1 & 1 & 0 \\ 1 & 1 & -1 \\ 1 & -1 & -z_2 \end{bmatrix} \begin{bmatrix} 1 \\ r \\ \Psi_{2a}' \end{bmatrix} = \begin{bmatrix} 1 & 1 & 0 \\ 0 & 0 & 1 \\ z_1 & -z_1 & -z_2 \end{bmatrix} \begin{bmatrix} \Psi_{1a} \\ \Psi_{1b} \\ \Psi_{2b}' \end{bmatrix}$$

$$\begin{bmatrix} 1 \\ r \\ \Psi_{2a}' \end{bmatrix} = \frac{1}{2}\begin{bmatrix} z_2 + z_1 + 1 & z_2 - z_1 + 1 & -2z_2 \\ -z_2 - z_1 + 1 & -z_2 + z_1 + 1 & 2z_2 \\ 2 & 2 & -2 \end{bmatrix} \begin{bmatrix} \Psi_{1a} \\ \Psi_{1b} \\ \Psi_{2b}' \end{bmatrix}$$

At the other end

$$\begin{bmatrix} 1 & 1 & 0 \\ 0 & 0 & -e^{j\phi_{2a}} \\ 1 & -1 & -z_2 e^{j\phi_{2a}} \end{bmatrix} \begin{bmatrix} t_R \\ t_L \\ \Psi_{2a}' \end{bmatrix} = \begin{bmatrix} 1 & 1 & 0 \\ 0 & 0 & 1 \\ z_1 & -z_1 & -z_2 \end{bmatrix} \begin{bmatrix} e^{j\phi_{1a}} & 0 & 0 \\ 0 & e^{-j\phi_{1b}} & 0 \\ 0 & 0 & e^{-j\phi_{2b}} \end{bmatrix} \begin{bmatrix} \Psi_{1a} \\ \Psi_{1b} \\ \Psi_{2b}' \end{bmatrix}$$

$$\begin{bmatrix} e^{-j\phi_{1a}} & 0 & 0 \\ 0 & e^{j\phi_{1b}} & 0 \\ 0 & 0 & e^{j\phi_{2a}} \end{bmatrix} \frac{1}{2z_1} \begin{bmatrix} z_2 + z_1 + 1 & z_2 + z_1 - 1 & -2z_2 e^{j\phi_{2a}} \\ -z_2 + z_1 - 1 & -z_2 + z_1 + 1 & 2z_2 e^{j\phi_{2a}} \\ 2 & 2 & -2e^{j\phi_{2a}} \end{bmatrix} \begin{bmatrix} t_R \\ t_L \\ \Psi_{2a}' \end{bmatrix} = \begin{bmatrix} \Psi_{1a} \\ \Psi_{1b} \\ \Psi_{2b}' \end{bmatrix}$$

Combining the equations at both ends:

$$\begin{bmatrix} 1 \\ r \\ \Psi_{2a}' \end{bmatrix} = \frac{1}{4z_1} \begin{bmatrix} z_2 + z_1 + 1 & z_2 - z_1 + 1 & -2z_2 \\ -z_2 - z_1 + 1 & -z_2 + z_1 + 1 & 2z_2 \\ 2 & 2 & -2 \end{bmatrix} \begin{bmatrix} e^{-j\phi_{1a}} & 0 & 0 \\ 0 & e^{j\phi_{1b}} & 0 \\ 0 & 0 & e^{j\phi_{2a}} \end{bmatrix} \begin{bmatrix} z_2 + z_1 + 1 & z_2 + z_1 - 1 & -2z_2 e^{j\phi_{2a}} \\ -z_2 + z_1 - 1 & -z_2 + z_1 + 1 & 2z_2 e^{j\phi_{2a}} \\ 2 & 2 & -2e^{j\phi_{2a}} \end{bmatrix} \begin{bmatrix} t_R \\ t_L \\ \Psi_{2a}' \end{bmatrix}$$

Here, we incorporate the fact that $z_1 = 1$ for our system to get

$$\begin{bmatrix} 1 \\ r \\ \Psi_{2a}' \end{bmatrix} = \frac{1}{4} \begin{bmatrix} 2 + z_2 & z_2 & -2z_2 \\ -z_2 & 2 - z_2 & 2z_2 \\ 2 & 2 & -2 \end{bmatrix} \begin{bmatrix} e^{-j\phi_{1a}} & 0 & 0 \\ 0 & e^{j\phi_{1b}} & 0 \\ 0 & 0 & e^{j\phi_{2b}} \end{bmatrix} \begin{bmatrix} 2 + z_2 & z_2 & -2z_2 e^{j\phi_{2a}} \\ -z_2 & 2 - z_2 & 2z_2 e^{j\phi_{2a}} \\ 2 & 2 & -2e^{j\phi_{2a}} \end{bmatrix} \begin{bmatrix} t_R \\ t_L \\ \Psi_{2a}' \end{bmatrix}$$

Or, written another way:

`

$$\begin{bmatrix} 1 \\ r \\ \Psi'_{2a} \end{bmatrix} = \begin{bmatrix} M_{Theta_{11}} & M_{Theta_{12}} & M_{Theta_{13}} \\ M_{Theta_{21}} & M_{Theta_{22}} & M_{Theta_{23}} \\ M_{Theta_{31}} & M_{Theta_{32}} & M_{Theta_{33}} \end{bmatrix} \begin{bmatrix} t_R \\ t_L \\ \Psi'_{2a} \end{bmatrix}$$

Where $\begin{bmatrix} M_{Theta_{11}} & M_{Theta_{12}} & M_{Theta_{13}} \\ M_{Theta_{21}} & M_{Theta_{22}} & M_{Theta_{23}} \\ M_{Theta_{31}} & M_{Theta_{32}} & M_{Theta_{33}} \end{bmatrix}$ is given by

$$\begin{bmatrix} M_{Theta_{11}} & M_{Theta_{12}} & M_{Theta_{13}} \\ M_{Theta_{21}} & M_{Theta_{22}} & M_{Theta_{23}} \\ M_{Theta_{31}} & M_{Theta_{32}} & M_{Theta_{33}} \end{bmatrix} = \frac{1}{4} \begin{bmatrix} z_2+2 & z_2 & -2z_2 \\ -z_2 & -z_2+2 & 2z_2 \\ 2 & 2 & -2 \end{bmatrix} \begin{bmatrix} e^{-j\phi_{1a}} & 0 & 0 \\ 0 & e^{j\phi_{1b}} & 0 \\ 0 & 0 & e^{j\phi_{2a}} \end{bmatrix} \begin{bmatrix} 2+z_2 & z_2 & -2z_2e^{j\phi_{2a}} \\ -z_2 & 2-z_2 & 2z_2e^{j\phi_{2a}} \\ 2 & 2 & -2e^{j\phi_{2a}} \end{bmatrix}$$

Solving for $\Psi'_{2a}$ in terms of the other variables using the third equation:

$$\Psi'_{2a} = \frac{M_{Theta_{31}}}{1 - M_{Theta_{33}}} t_R + \frac{M_{Theta_{32}}}{1 - M_{Theta_{33}}} t_L$$

Plugging this back into the equation above:

$$\begin{bmatrix} 1 \\ r \end{bmatrix} = \begin{bmatrix} M_{Theta_{11}} & M_{Theta_{12}} & M_{Theta_{13}} \\ M_{Theta_{21}} & M_{Theta_{22}} & M_{Theta_{23}} \end{bmatrix} \begin{bmatrix} 1 & 0 \\ 0 & 1 \\ \frac{M_{Theta_{31}}}{1 - M_{Theta_{33}}} & \frac{M_{Theta_{32}}}{1 - M_{Theta_{33}}} \end{bmatrix} \begin{bmatrix} t_R \\ t_L \end{bmatrix}$$

$$\begin{bmatrix} 1 \\ r \end{bmatrix} = \begin{bmatrix} M_{Theta_{11}} + \frac{M_{Theta_{13}} M_{Theta_{31}}}{1 - M_{Theta_{33}}} & M_{Theta_{12}} + \frac{M_{Theta_{13}} M_{Theta_{32}}}{1 - M_{Theta_{33}}} \\ M_{Theta_{21}} + \frac{M_{Theta_{23}} M_{Theta_{31}}}{1 - M_{Theta_{33}}} & M_{Theta_{22}} + \frac{M_{Theta_{23}} M_{Theta_{32}}}{1 - M_{Theta_{33}}} \end{bmatrix} \begin{bmatrix} t_R \\ t_L \end{bmatrix}$$

Since, for the single-ring theta cavity, there is no light entering from the right, $t_L = 0$. Therefore:

$$t = t_R = \frac{1 - M_{Theta_{33}}}{M_{Theta_{11}}(1 - M_{Theta_{33}}) + M_{Theta_{13}} M_{Theta_{31}}}$$

$$r = \frac{M_{Theta_{21}}(1 - M_{Theta_{33}}) + M_{Theta_{23}} M_{Theta_{31}}}{M_{Theta_{11}}(1 - M_{Theta_{33}}) + M_{Theta_{13}} M_{Theta_{31}}}$$

Plugging in the equations for the $M_{Theta}$ matrix above, these equations become:

$$t(\phi_1, \phi_2) = \frac{2z(z sin(\phi_1) + sin(\phi_2))}{[2z(2\, cos(\phi_2) - e^{j\phi_1}) - j(z^2+4) sin(\phi_2)](z\, sin(\phi_1) + sin(\phi_2)) - j(z(cos(\phi_2) - e^{j\phi_1}) - 2j\, sin(\phi_2))^2}$$

$$r(\phi_1, \phi_2) = \frac{(z\, sin(\phi_1) + sin(\phi_2))(2z(e^{j\phi_1} - \cos(\phi_2)) + jz^2 \sin(\phi_2)) + j\left(z^2(\cos(\phi_2) - e^{j\phi_1})^2 - 2jz \sin(\phi_2)(cos(\phi_2) - e^{j\phi_1})\right)}{[2z(2\, cos(\phi_2) - e^{j\phi_1}) - j(z^2+4) sin(\phi_2)](z\, sin(\phi_1) + sin(\phi_2)) - j(z(cos(\phi_2) - e^{j\phi_1}) - 2j\, sin(\phi_2))^2}$$

Where we have assumed the following in addition to the assumptions discussed above.

a) based on reciprocity $\boldsymbol{\phi_{1a} = \phi_{1b}, \phi_{2a} = \phi_{2b}}$
b) Based on the fact that all waveguides are made of the same material and have the same width: $z_1 = 1$
c) We also let $z_2 = z$ since it is the only effective coupling term not equal to 1.

`

Using different parameters, $\phi_2 = \phi$, $\phi_1 = \gamma\phi$, where $\boldsymbol{\gamma} = \frac{\boldsymbol{L_1}}{\boldsymbol{L_2}}$, we get

$$r(\phi) = \frac{(zsin(\gamma\phi) + sin(\phi))(-2z(cos(\phi) - e^{j\gamma\phi}) + jz^2 sin(\phi)) + j((zcos(\phi) - ze^{j\gamma\phi})^2 - 2jzsin(\phi))}{(zsin(\gamma\phi) + sin(\phi))(4zcos(\phi) - j(z^2+4)\,sin(\phi) - 2ze^{j\gamma\phi}) - j(z(cos(\phi) - e^{j\gamma\phi}) - 2j\,sin(\phi))^2}$$

$$t(\phi) = \frac{2z(zsin(\gamma\phi) + sin(\phi))}{(zsin(\gamma\phi) + sin(\phi))(4zcos(\phi) - j(z^2+4)\,sin(\phi) - 2ze^{j\gamma\phi}) - j(z(cos(\phi) - e^{j\gamma\phi}) - 2j\,sin(\phi))^2}$$

$$\Psi_2 = \frac{(1+r)e^{-j\phi}}{2j\,sin(\phi)} - t$$

Note that in estimating $\gamma$, we are assuming the waveguides have the same effective index of refraction. By taking the magnitude of the wavefunctions we obtain the spectra for reflection, transmission, and Electric field enhancement in ring resonator (path 2). The phase $\phi$, can be related to the expected mode number, by considering the effective path length in the ring resonator, described in equation below:

$$\frac{2L_2}{\lambda_{eff}} = \frac{\phi}{\pi} = \boldsymbol{M} \equiv \textit{ mode \#}$$

The raw spectra obtained from the wavefunctions is shown in Fig S2. The manifolds are created by extracting mode characteristics as a function of phase and mapping their behavior with changing design parameters, shown in Fig S3.

### 1.2 Nested Theta Cavity Analytical Model Derivations

Next, we detail the derivation of the analytical wave equations for the Nested Theta Cavity using the physical model shown in Fig S10. To begin, we start by defining the interface-continuity equations for the MSC junction at the entrance port:

$$1 + r = \Psi_{1l,a} + \Psi_{1l,b}$$
$$1 + r = \Psi'_{out,a} + \Psi'_{out,b}$$
$$1 - r = (\Psi_{1l,a} - \Psi_{1l,b}) + z_2(\Psi'_{out,a} - \Psi'_{out,b})$$

And the MSC junction at the exit gives us:

$$t = e^{j\phi_{1r}}\Psi_{1r,a} + e^{-j\phi_{1r}}\Psi_{1l,b}$$
$$t = e^{j\phi_o}\Psi'_{out,a} + e^{-j\phi_o}\Psi'_{out,b}$$
$$t = (e^{j\phi_{1r}}\Psi_{1r,a} - e^{-j\phi_{1r}}\Psi_{1r,b}) + z_2(e^{j\phi_o}\Psi'_{out,a} - e^{-j\phi_o}\Psi'_{out,b})$$

Where we are assuming the interfaces/junctions all have the same dimensions and effective parameters. Using the same approach as for the single ring Theta Cavity, these equations become:

$$\begin{bmatrix} 1 \\ r \\ \Psi'_{out,a} \end{bmatrix} = \begin{bmatrix} 1 & 1 & 0 \\ 0 & 0 & -1 \\ 1 & -1 & -z_2 \end{bmatrix}^{-1} \begin{bmatrix} 1 & 1 & 0 \\ 0 & 0 & 1 \\ 1 & -1 & -z_2 \end{bmatrix} \begin{bmatrix} \Psi_{11,a} \\ \Psi_{11,b} \\ \Psi'_{out,b} \end{bmatrix}$$

`

$$\begin{bmatrix}1 & 1 & 0\\ 0 & 0 & 1\\ 1 & -1 & -z_2\end{bmatrix}^{-1}\begin{bmatrix}1 & 1 & 0\\ 0 & 0 & -e^{j\phi_o}\\ 1 & -1 & -z_2e^{-j\phi_o}\end{bmatrix}\begin{bmatrix}t\\ 0\\ \Psi'_{out,a}\end{bmatrix}=\begin{bmatrix}e^{j\phi_{1l}}\Psi_{1r,a}\\ e^{-j\phi_{1l}}\Psi_{1r,b}\\ e^{-j\phi_o}\Psi'_{out,b}\end{bmatrix}$$

So it would be convenient to relate $\Psi$ parameters here, i.e.

$$\begin{bmatrix}e^{j\phi_{1l}}\Psi_{1r,a}\\ e^{-j\phi_{1l}}\Psi_{1r,b}\end{bmatrix}=\begin{bmatrix}T_{11} & T_{12}\\ T_{21} & T_{22}\end{bmatrix}\begin{bmatrix}\Psi_{1l,a}\\ \Psi_{1l,b}\end{bmatrix}$$

So that we would get

$$\begin{bmatrix}1\\ r\\ \Psi'_{out,a}\end{bmatrix}=\frac{1}{4}\begin{bmatrix}2+z_2 & z_2 & -2z_2\\ -z_2 & 2-z_2 & 2z_2\\ 2 & 2 & -2\end{bmatrix}\begin{bmatrix}T_{11} & T_{12} & 0\\ T_{21} & T_{22} & 0\\ 0 & 0 & -e^{-j\phi_o}\end{bmatrix}^{-1}\begin{bmatrix}2+z_2 & z_2 & -2z_2e^{j\phi_o}\\ -z_2 & 2-z_2 & 2z_2e^{j\phi_o}\\ 2 & 2 & -2e^{j\phi_o}\end{bmatrix}\begin{bmatrix}t\\ 0\\ \Psi'_{out,a}\end{bmatrix}$$

So that we could combine our system to equations to get

$$\begin{bmatrix}1\\ r\\ \Psi'_{out,a}\end{bmatrix}=[M_{Total}]\begin{bmatrix}t\\ 0\\ \Psi'_{out,a}\end{bmatrix}$$

$$[M_{Total}]=\frac{1}{4}\begin{bmatrix}2+z_2 & z_2 & -2z_2\\ -z_2 & 2-z_2 & 2z_2\\ 2 & 2 & -2\end{bmatrix}\begin{bmatrix}T_{11} & T_{12} & 0\\ T_{21} & T_{22} & 0\\ 0 & 0 & -e^{-j\phi_o}\end{bmatrix}^{-1}\begin{bmatrix}2+z_2 & z_2 & -2z_2e^{j\phi_o}\\ -z_2 & 2-z_2 & 2z_2e^{j\phi_o}\\ 2 & 2 & -2e^{j\phi_o}\end{bmatrix}\begin{bmatrix}t\\ 0\\ \Psi'_{out,a}\end{bmatrix}$$

$$[M_{Total}]=\frac{1}{4}\begin{bmatrix}z_2+2 & z_2 & -2z_2\\ -z_2 & -z_2+2 & 2z_2\\ 2 & 2 & -2\end{bmatrix}\begin{bmatrix}\frac{T_{22}}{T_{11}T_{22}-T_{21}T_{12}} & \frac{-T_{12}}{T_{11}T_{22}-T_{21}T_{12}} & 0\\ \frac{-T_{21}}{T_{11}T_{22}-T_{21}T_{12}} & \frac{T_{11}}{T_{11}T_{22}-T_{21}T_{12}} & 0\\ 0 & 0 & -e^{j\phi_o}\end{bmatrix}\begin{bmatrix}2+z_2 & z_2 & -2z_2e^{j\phi_o}\\ -z_2 & 2-z_2 & 2z_2e^{j\phi_o}\\ 2 & 2 & -2e^{j\phi_o}\end{bmatrix}$$

We therefore need only solve for the $[T_1]$ matrix. Using the same approach, we used the single ring theta cavity, and substituting the terms relevant to this situation, we find:

$$\begin{bmatrix}e^{j\phi_{1l}}\Psi_{1l,a}\\ e^{-j\phi_{1l}}\Psi_{1l,b}\end{bmatrix}=\begin{bmatrix}M_{inner_{11}}+\frac{M_{inner_{13}}M_{inner_{31}}}{1-M_{inner_{33}}} & M_{inner_{12}}+\frac{M_{inner_{13}}M_{inner_{32}}}{1-M_{inner_{33}}}\\ M_{inner_{21}}+\frac{M_{inner_{23}}M_{inner_{31}}}{1-M_{inner_{33}}} & M_{inner_{22}}+\frac{M_{inner_{23}}M_{inner_{32}}}{1-M_{inner_{33}}}\end{bmatrix}\begin{bmatrix}\Psi_{1r,a}\\ \Psi_{1r,b}\end{bmatrix}$$

Assuming $\phi_{inner}=\phi_i$, the transfer matrix can be solved as:

$$M_{inner}:=\frac{1}{4}\begin{bmatrix}z_2+2 & z_2 & -2z_2\\ -z_2 & -z_2+2 & 2z_2\\ 2 & 2 & -2\end{bmatrix}\begin{bmatrix}e^{-j\phi_1} & 0 & 0\\ 0 & e^{j\phi_1} & 0\\ 0 & 0 & e^{j\phi_i}\end{bmatrix}\begin{bmatrix}2+z_2 & z_2 & -2z_2e^{j\phi_i}\\ -z_2 & 2-z_2 & 2z_2e^{j\phi_i}\\ 2 & 2 & -2e^{j\phi_i}\end{bmatrix}$$

Which means that

$$\begin{bmatrix}e^{j\phi_{1r}}\Psi_{1r,a}\\ e^{-j\phi_{1r}}\Psi_{1r,b}\end{bmatrix}=\begin{bmatrix}e^{j\phi_{1r}} & 0\\ 0 & e^{-j\phi_{1r}}\end{bmatrix}\begin{bmatrix}M_{inner_{11}}+\frac{M_{inner_{13}}M_{inner_{31}}}{1-M_{inner_{33}}} & M_{inner_{12}}+\frac{M_{inner_{13}}M_{inner_{32}}}{1-M_{inner_{33}}}\\ M_{inner_{21}}+\frac{M_{inner_{23}}M_{inner_{31}}}{1-M_{inner_{33}}} & M_{inner_{22}}+\frac{M_{inner_{23}}M_{inner_{32}}}{1-M_{inner_{33}}}\end{bmatrix}^{-1}\begin{bmatrix}e^{j\phi_{1l}} & 0\\ 0 & e^{-j\phi_{1l}}\end{bmatrix}\begin{bmatrix}\Psi_{1l,a}\\ \Psi_{1l,b}\end{bmatrix}$$

$$\begin{bmatrix}T_{11} & T_{12}\\ T_{21} & T_{22}\end{bmatrix}=\begin{bmatrix}e^{j\phi_{1r}} & 0\\ 0 & e^{-j\phi_{1r}}\end{bmatrix}\begin{bmatrix}M_{inner_{11}}+\frac{M_{inner_{13}}M_{inner_{31}}}{1-M_{inner_{33}}} & M_{inner_{12}}+\frac{M_{inner_{13}}M_{inner_{32}}}{1-M_{inner_{33}}}\\ M_{inner_{21}}+\frac{M_{inner_{23}}M_{inner_{31}}}{1-M_{inner_{33}}} & M_{inner_{22}}+\frac{M_{inner_{23}}M_{inner_{32}}}{1-M_{inner_{33}}}\end{bmatrix}^{-1}\begin{bmatrix}e^{j\phi_{1l}} & 0\\ 0 & e^{-j\phi_{1l}}\end{bmatrix}$$

$$\begin{bmatrix}T_{11} & T_{12}\\ T_{21} & T_{22}\end{bmatrix}^{-1}=\begin{bmatrix}e^{-j\phi_{1l}} & 0\\ 0 & e^{j\phi_{1l}}\end{bmatrix}\begin{bmatrix}M_{inner_{11}}+\frac{M_{inner_{13}}M_{inner_{31}}}{1-M_{inner_{33}}} & M_{inner_{12}}+\frac{M_{inner_{13}}M_{inner_{32}}}{1-M_{inner_{33}}}\\ M_{inner_{21}}+\frac{M_{inner_{23}}M_{inner_{31}}}{1-M_{inner_{33}}} & M_{inner_{22}}+\frac{M_{inner_{23}}M_{inner_{32}}}{1-M_{inner_{33}}}\end{bmatrix}\begin{bmatrix}e^{-j\phi_{1r}} & 0\\ 0 & e^{j\phi_{1r}}\end{bmatrix}$$

`

Which is pertinent as

$$\begin{bmatrix} T_{11} & T_{12} \\ T_{21} & T_{22} \end{bmatrix}^{-1} = \begin{bmatrix} \frac{T_{22}}{T_{11}T_{22} - T_{21}T_{12}} & \frac{-T_{12}}{T_{11}T_{22} - T_{21}T_{12}} \\ \frac{-T_{21}}{T_{11}T_{22} - T_{21}T_{12}} & \frac{T_{11}}{T_{11}T_{22} - T_{21}T_{12}} \end{bmatrix}$$

These equations can now be combined to yield the expression for spectra, $r(\phi)$ and $t(\phi)$. Similarly, we can solve for $\Psi'_{out,a}$ and $\Psi'_{out,b}$ and normalize to get:

$$t = \frac{1 - M_{Total_{33}}}{M_{Total_{11}}(1 - M_{Total_{33}}) + M_{Total_{13}} M_{Total_{31}}}$$

$$r = \frac{M_{Total_{21}}(1 - M_{Total_{33}}) + M_{Total_{23}} M_{Total_{31}}}{M_{Total_{11}}(1 - M_{Total_{33}}) + M_{Total_{13}} M_{Total_{31}}}$$

$$\Psi_{out,a} = \sqrt{\frac{2}{z_2}} * \frac{M_{Total_{31}}}{M_{Total_{11}}(1 - M_{Total_{33}}) + M_{Total_{13}} M_{Total_{31}}}$$

$$\Psi_{out,b} = \sqrt{\frac{2}{z_2}} * \left(1 + \frac{M_{total_{21}}(1 - M_{Total_{33}}) + (M_{Total_{23}} - 1) M_{Total_{31}}}{M_{Total_{11}}(1 - M_{Total_{33}}) + M_{Total_{13}} M_{Total_{31}}}\right)$$

Similarly, one can solve for the equivalent terms in the inner ring, $\Psi_{in,a}$ and $\Psi_{in,b}$ by solving for the remaining outer terms in terms of $r$, $t$, $\Psi_{out,a}$, and$\Psi_{out,b}$then setting up and solving the remaining system of equations. This comes out to:

$$\Psi_{in,a} = \frac{\sqrt{\frac{2}{z_2}} e^{j\left(\frac{\phi_1}{2} - \phi_{in}\right)} + z_2 \sin(\phi_1)\left(\Psi_{out,b}(e^{-j\phi_{in}} + e^{-j\phi_o}) - \Psi_{out,a}(e^{-j\phi_i} + e^{j\phi_o})\right) + r\sqrt{\frac{2}{z_2}} e^{j\left(\frac{\phi_1}{2} + \phi_i\right)} - t\sqrt{\frac{2}{z_2}} e^{-\frac{j\phi_1}{2}}}{-\mathrm{j}\sin(\phi_i)}$$

$$\Psi_{in,b} = \frac{\sqrt{\frac{2}{z_2}} e^{j\left(\frac{\phi_1}{2} + \phi_i\right)} + z_2 \sin(\phi_1)\left(\Psi_{out,b}(e^{j\phi_i} + e^{-j\phi_o}) - \Psi_{out,a}(e^{j\phi_i} + e^{j\phi_o})\right) + r\sqrt{\frac{2}{z_2}} e^{j\left(\frac{\phi_1}{2} - \phi_i\right)} - t\sqrt{\frac{2}{z_2}} e^{-j\frac{\phi_1}{2}}}{\mathrm{j}\sin(\phi_i)}$$

After obtaining the equations for the wavefunctions as a general function of the relative phases in each path, the spectra can be derived by taking the magnitude of the wavefunction. A reduction of variables into a single-phase value is done by including design constraints based on the relative phase relationship set by the nest ring geometry. For our design, the design constraints are defined below:

$$\phi_1 = \frac{\gamma\phi}{2} + j\delta$$

$$\phi_{11} = \phi_{1l} = \phi_{1r} = \frac{\gamma\phi}{4} + j\delta$$

$$\phi_{inner} = \frac{\gamma\phi}{2(\gamma + \Delta)} + j\delta$$

$$\phi_{outer} = \frac{\gamma\phi}{\gamma + \Delta} + j\delta$$

Notice the inclusion of loss is assumed to be equivalent for all paths. Also, the change in gamma ($\Delta$) is plotted along the x-axis in the dispersion relation; for the dynamic outer ring case, $\Delta$=0 for $\phi_{inner}$ and the dynamic inner ring case $\Delta$=0 for $\phi_{outer}$.

`

## 2. Experimental

The analytical model demonstrates strong agreement with simulation and experimental measurements, particularly in predicting the behavior of qBIC and BIC resonant modes and the associated spectral patterns. In experimental devices, these modes can deviate from idealized predictions, highlighting the influence of physical effects that are not captured by the analytical framework. In particular, the model and simulations assume ideal single mode operation, perfect structural symmetry, and loss mechanisms that are uniform and well behaved. Although non-idealities in real devices don't preclude the physics predicted by the models, we find several non-idealities can contribute to discrepancies which are examined in detail in this supplemental section.

### 2.1 Fabrication Tolerances and Effective Medium Approximation

The spectral response of the Theta Cavity is governed by the gamma coefficient, which is directly determined by the physical geometry of the device. This dependence enables engineering of transmission characteristics through careful control of cavity dimensions and junction geometry. In this work, we experimentally demonstrate the TC on a Silicon on Insulator platform operating at telecom wavelengths near 1550 nm. Devices are fabricated using a standard single layer process with fully etched single mode waveguides of 500 nm width and 220 nm height, together with tapered grating couplers for optical access.

In practice, fabrication imperfections can introduce slight deviations from the idealized geometry assumed in analytical and simulation models. Variations such as uneven etching, side wall and line width roughness, or imperfect feature sizes can locally modify the effective refractive index of the waveguide. Fig S6 highlights several aspects of these fabrication non-idealities. Scanning electron microscope images of a circular theta cavity illustrate the overall device geometry (Fig S11a), while a magnified view of the MSC junction (Fig S11c) emphasizes the symmetry required to realize the desired gamma coefficient. The effect of effective index is illustrated in Fig S11b by showing the relationship between waveguide effective index and free spectral range as a function of ring diameter. As shown, higher effective index theta cavities exhibit a reduced free spectral range. This trade off underscores the importance of fabrication precision and motivates the use of effective medium approximations as a practical, though imperfect, description of real devices.

Fig S6a illustrates how fabrication non-idealities in the theta cavity geometry affect the mode characteristics. We quantify them as such: waveguide offset ($\Delta y$), change in waveguide thickness ($\Delta w$), and radius of curvature ($\Delta r$) within the MSC junction. Fig S6b shows that though we introduce imperfections through increases in parameters $\Delta y$, $\Delta w$, and $\Delta r$, the mode behavior is preserved across our simulation bandwidth with only slight changes in quality factor. These results suggest that the topological response of the theta cavity is robust to fabrication non-idealities within the expected limits of modern electron beam lithography and SOI photonic processing.

### 2.2 Multimode Waveguide Effects

While the analytical and numerical models assume single mode waveguide operation, practical Theta Cavity devices fabricated using a standard Silicon on Insulator process can support higher order modes over certain wavelength ranges. This limitation is inherent to the waveguide cross sections accessible within a single layer SOI platform and becomes more pronounced as waveguide widths increase.

The higher order waveguide modes can alter the observed transmission spectrum. Multiple resonant modes can spectrally overlap and form what may be described as "bimodal"

`

resonances with multiple peaks. These; these overlapping modes can (TE10 and TE11) can interact with one another producing unique spectral features. Capturing the effects of higher order waveguide modes in the analytical framework which targets coupling of the model require the inclusion of a model parameters (z, γ, δ) unique to the higher order mode (TE11) due to their difference in mode propagation characteristics compared to the fundamental waveguide mode (TE10). These effects are illustrated in Fig S12 through simulated and experimental results for circular theta cavities with a fixed diameter of 24.5 µm and increasing ring waveguide widths. Simulations comparing a 200 nm wide ring (Fig S12a and Fig S12b) and a 300 nm wide ring (Fig S12c and Fig S12d) show that increasing the waveguide width enables the emergence of higher order TC modes, as evidenced by the visualized electric field profiles and the increased number of resonance features in the transmission spectra. Experimental measurements from a device with a 500 nm wide ring further demonstrate this behavior, exhibiting ultradeep resonances where multiple modes overlap in wavelength. The mutual interaction of these modes alters the spectral response, leading to spectra patterns not predicted in the analytical model as presented in this study.

### 2.3 Geometric Design Tradeoffs

The gamma coefficient in a Theta Cavity, which is determined by the ratio of the middle waveguide path and ring resonator, is subject to design constraints that arise from waveguide routing, near-field coupling, and propagation loss. Fig S13 summarizes our design equations and parameters across three representative cavity configurations, shown in Fig 6, that target different gamma regimes, highlighting the challenges associated with designing target path length ratios in a real photonic TC devices.

For the balanced Theta Cavity configuration, where $\gamma \approx 1$, the inner waveguide path and the outer ring resonator are designed to have nearly identical effective optical path lengths. As shown schematically in Fig S13a, this is accomplished by introducing a U bend in the inner path to increase its length. The U-bend consists of four 90$^{o}$ bends, two vertical buffer sections of length $L_y$, and two straight buffer sections of length $L_x$ on each side. A key limitation in this design is the requirement to maintain a waveguide separation greater than 2.5 µm between the top of the U bend in the middle path and the ring resonator path to avoid unintended crosstalk from near-field coupling. This spacing constraint restricts how tightly the geometry can be folded, making it challenging to achieve a perfect $\gamma = 1$ condition in circular theta cavities with smaller diameters. In addition, small bend radii must be avoided to prevent excessive attenuation leading to mode broadening. Our TC geometry, shown in the SEM image of Fig 6a, has an approximate gamma value of $\gamma = 0.976$, considering the tradeoffs listed above.

In the cyclical Theta Cavity regime, where $\gamma \approx 1/3$, the inner waveguide path is designed to be approximately one third the length of the outer ring, as illustrated in Fig S13b. To fine tune the path length ratio without lengthening the inner path, a straight extension parameter Ly is introduced along the outer ring. Our TC geometry, shown in the SEM image of Fig 6d, adopts a pill shaped geometry and realizes an approximate gamma value of $\gamma = 0.332$.

The binary Theta Cavity regime corresponds to gamma values approaching zero, where the inner waveguide path is much shorter than the outer ring resonator, as depicted in Fig S13c. A true $\gamma = 0$ condition is not physically realizable since both paths must have a finite, non-zero path length. This is particularly challenging because path 1 must be made as short as possible, while remaining limited by minimum gaps to avoid crosstalk from near-field coupling, and path 2 must be made long enough without incurring excessive propagation loss and ultranarrow free spectral range. In our implementation, path 1 is reduced to a minimum length of 2.5 µm in accordance with cross talk design constraints. Our TC geometry, shown in the SEM image of Fig 6g, adopts an “hourglass” shape and achieves an approximate gamma value of $\gamma = 0.035$.

`

### 3. Quality Factor Extraction

The mode quality factors and bandwidth trends shown throughout this study were extracted using various methods described below. The algorithms used for data analysis were developed from open-source Python libraries. All spectra were analyzed using a base-10 logarithmic representation and plotted on a decibel (dB) scale.

For the analytical and simulation data, the extracted bandwidth was used to calculate the quality factor following established convention, where the full width at half maximum (FWHM) was determined using a fixed threshold of −3 dB. This extraction was straightforward because the analytical and simulated spectra closely approximated ideal behavior, with all modes exhibiting nearly perfectly symmetric Lorentzian line shapes and the background was inherently flat, free of noise or interference fringes.

For the measured data, the spectra exhibited asymmetric Fano line shapes, interference fringes arising from back-reflections within the photonic circuit, and measurement noise, all of which complicated bandwidth extraction. To address this, we implemented a standard numerical fitting algorithm that employed iterative optimization to estimate the fitting parameters of a Fano line shape for each mode in the spectra. The quality factor was then determined by extracting the bandwidth parameter from the fitted Fano profile. This method was applied to all modes in the measured spectra to estimate the maximum observed quality factor within each dataset. The results from the fitting method are presented in Fig S14, where the highest quality factor modes are shown for the spectra highlighted in the manuscript. Specifically, Fig S14a corresponds to the spectra in Fig 5b; Fig S14b corresponds to the spectra in Fig 6c and Fig S5b; Fig S14c corresponds to the spectra in Fig 6f and Fig S5c; Fig S14d corresponds to the spectra in Fig 6i and Fig S5d; and Fig S14e corresponds to the spectra in Fig 7c. This method was also applied to extract the maximum quality factors of the modes shown in Fig S7.

The bandwidth trends were also extracted from the measured spectra for direct comparison with the spectral patterns predicted by the analytical model. In contrast to the previously described Fano fitting approach, which was optimized for accurately estimating the quality factor of individual modes, extracting global bandwidth trends required a more robust method that could be consistently applied across all measured modes. The primary challenge was that the fitting algorithm did not perform reliably for every mode due to background noise and distortion of the measured line shapes. In some cases, the iterative fitting routine either failed to converge or produced poor fits to the mode profile, resulting in missing data points that obscured the underlying pattern of the spectral bandwidth trends. To address this limitation, the background variation was subtracted to flatten measured spectra, after which the mode bandwidth was evaluated using a fixed threshold of −5 dB. This threshold was selected to ensure that the extracted bandwidth trends were not directly influenced by oscillations of the background noise, measured to be approximately 2 dB. Although the flat-threshold method does not provide the same level of accuracy as Fano fitting method for determining the FWHM of the individual modes, it significantly improved the consistency and precision of the spectral pattern observed in extracted bandwidth trends by minimizing sensitivity to variations in mode shape and background noise. This approach was therefore used to extract the bandwidth trends presented in Fig 5c, Fig S4, and Fig S5.

### Data availability

The data that support the plots within this paper and other findings of this study are available from the corresponding author upon reasonable request.

`

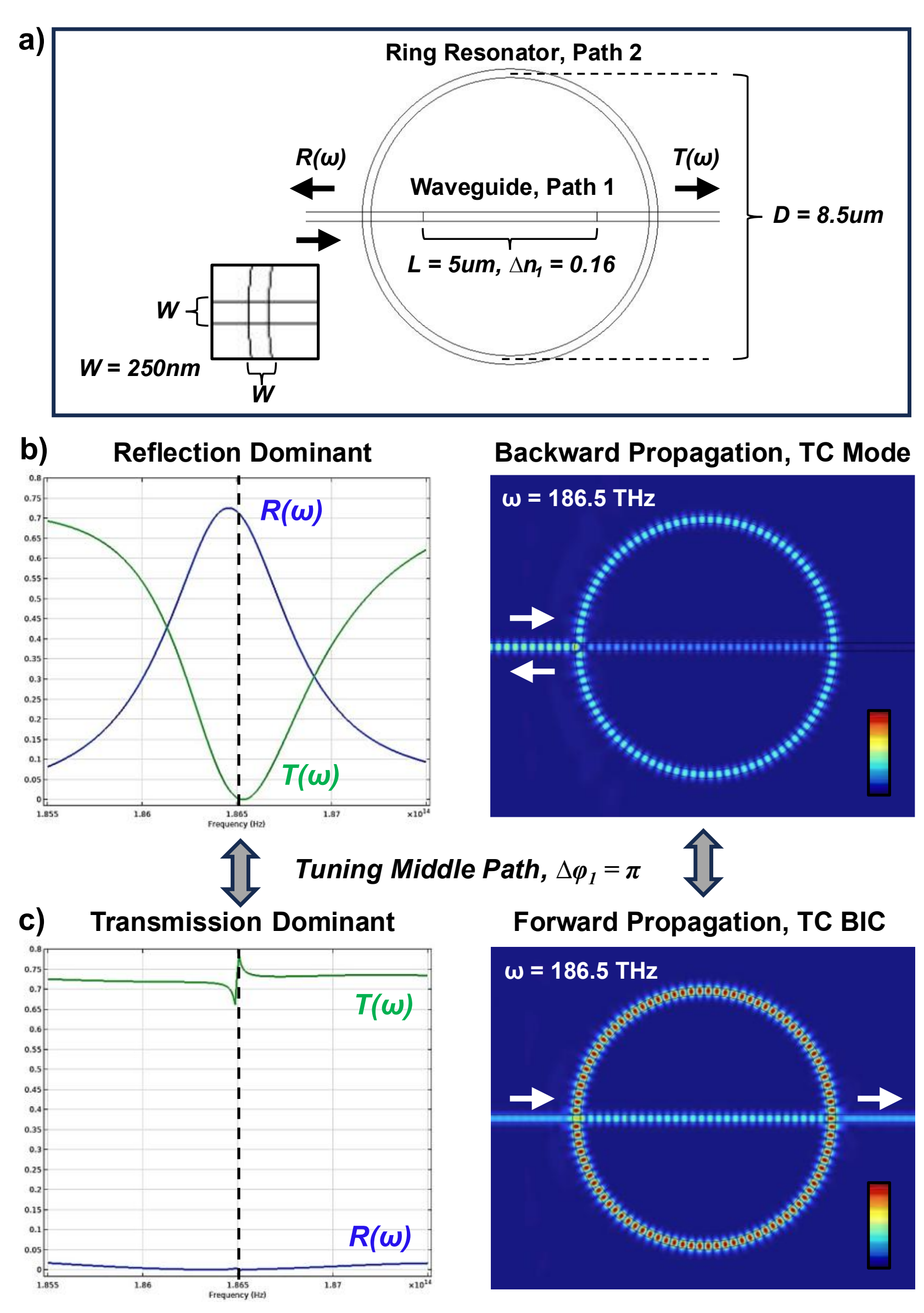

**Fig. S1. Amplitude Tuning.** (a) Model used to simulate Circular Theta Cavity (D = 8.5um, W = 250nm, n = 3.5) with the tuning of refractive index in segment of middle path to achieve $\Delta\varphi_1 = \pi$ to modulate the amplitude of the observable resonance. The spectra and normalized electric field is shown for a reflection dominate resonance (b) and a transmission dominate resonance (c) at a fixed frequency, 186.5 THz.

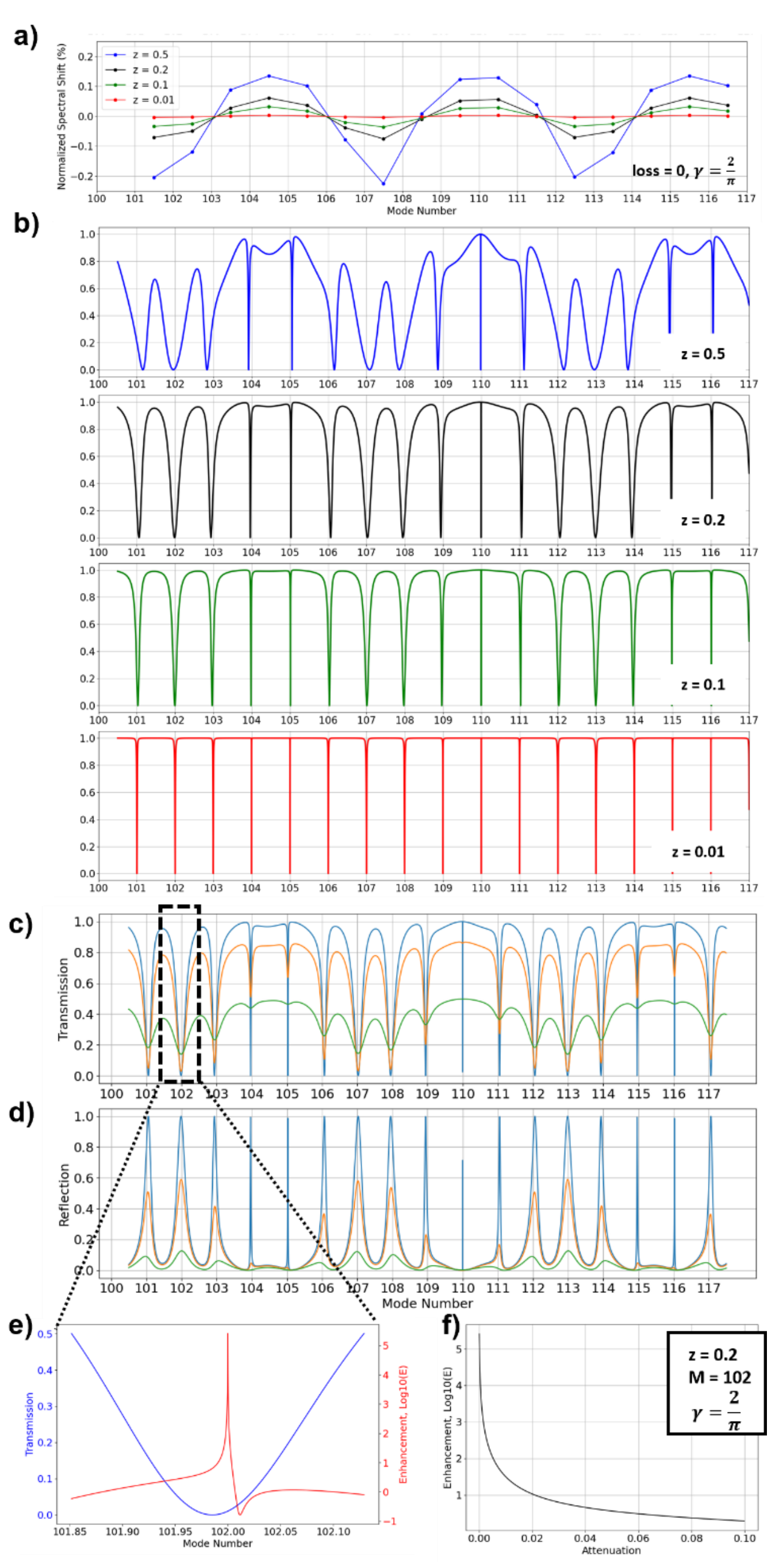

**Fig S2: Analytical Model Spectra.** (a) The spectral shift for increasing M-number is extracted from analytical transmission spectra plotted with larger z-coefficients resulting in larger shifts. (b) Predicted transmission spectra for circular Theta Cavity, loss = 0 and $\gamma = \frac{2}{\pi}$, with reducing coupling, z-coefficient shown from top to bottom. The analytical transmission (c) and reflection spectra (d) is plotted with different loss coefficients with larger values resulting in reduced peak height. The predicted line shape for transmission and electric field enhancement is shown with loss δ = 0.0001 (d) and the enhancement versus loss coefficient is plotted (e) for a representative mode (M=102, z = 0.2).

`

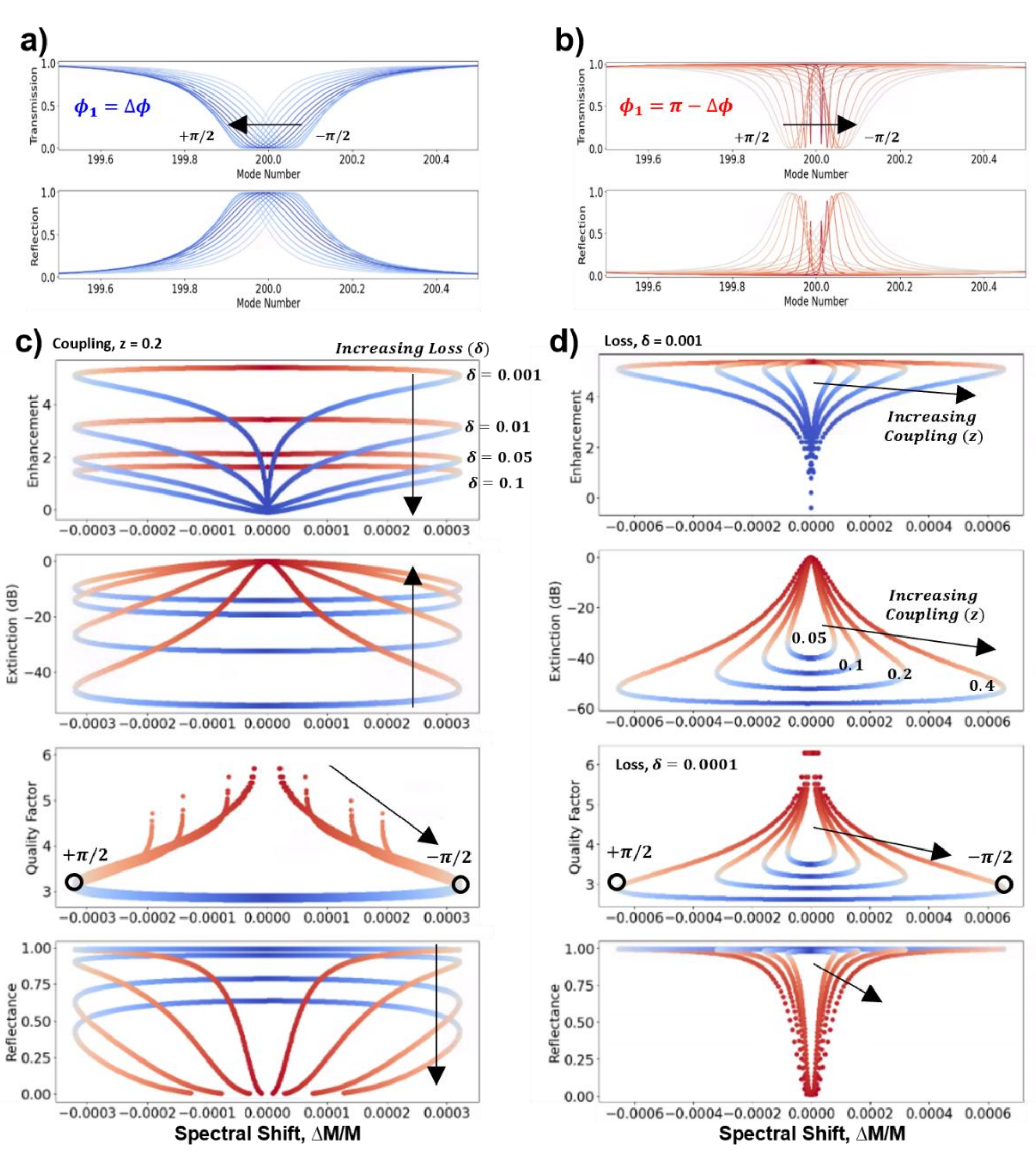

**Fig S3: Analytical Phase Manifolds.** The numerical analysis produces the single TC mode shape as the path 1 phase, $\varphi_1$, is tuned from 0 to 2π. The transmission (upper plot) and reflection (lower plot) spectra for broad modes with an odd phase relationship (a, false color blue) and sharp modes with an even phase relationship (b, false color red). Extracting the normalized Electric Field enhancement, extinction in dB, quality factor, and reflectance for modes with changing phase 1 phase reveals the phase manifold predicted by the mode. (c) The plotted manifolds are overlayed with increasing z-coupling coefficient, z = 0.05 to z = 0.4 (denoted by arrow) to reveal cavity design tradeoffs. (d) The plotted manifolds are overlayed with increasing loss, δ = 0.001 to δ = 0.1 (denoted by arrow) to reveal non-Hermitian properties.

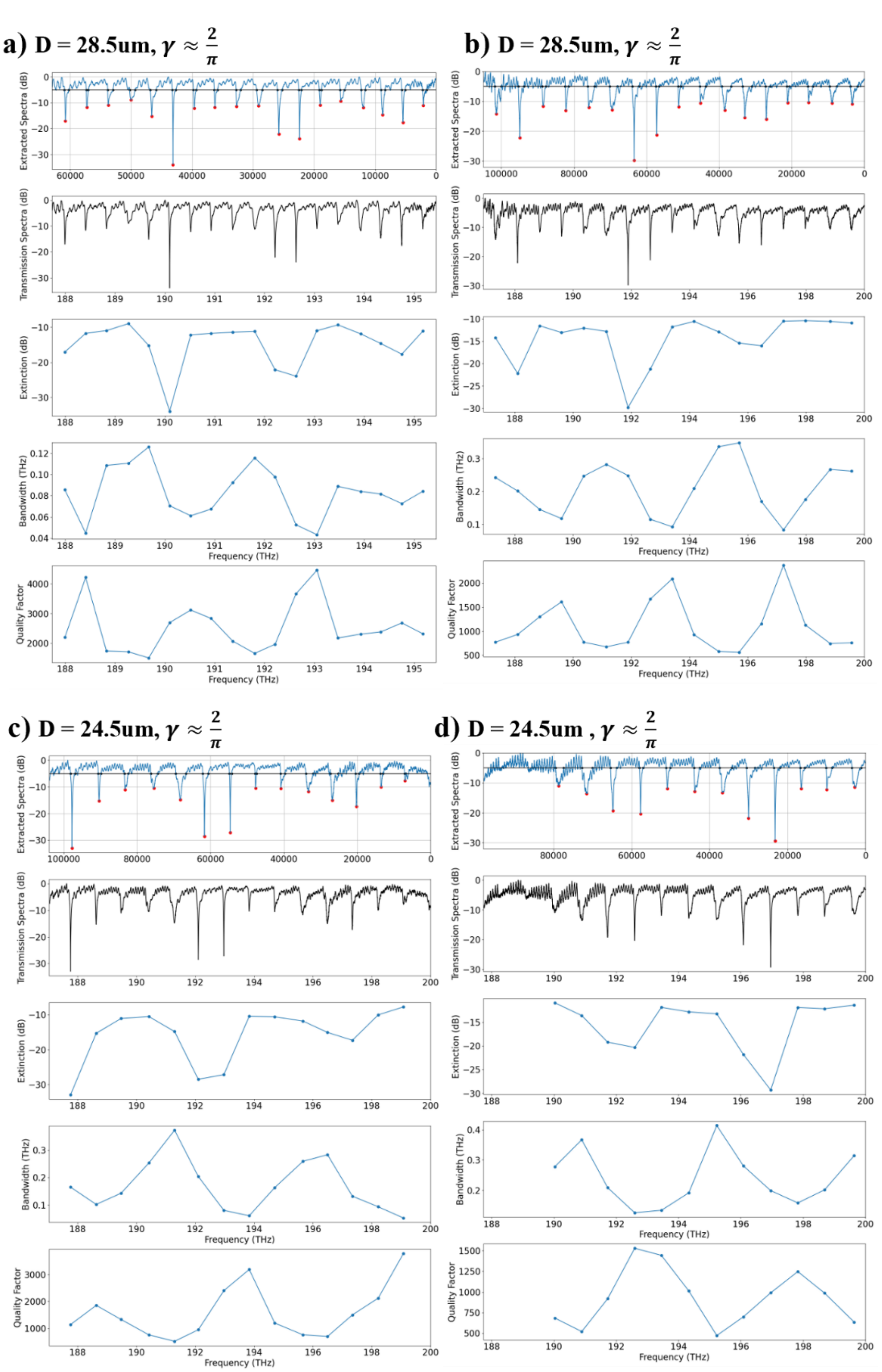

**Fig. S4. Measured Circular Theta Cavity Spectral Analysis.** Extracted spectra, measured transmission, extinction, bandwidth, and quality factor for circular theta cavities of diameters a) D = 28.5 µm with $\gamma \approx \frac{2}{\pi}$ , b) D = 28.5 µm with $\gamma \approx \frac{2}{\pi}$ , c) 24.5 µm with $\gamma \approx \frac{2}{\pi}$, and d) 24.5 µm with $\gamma \approx \frac{2}{\pi}$.

`

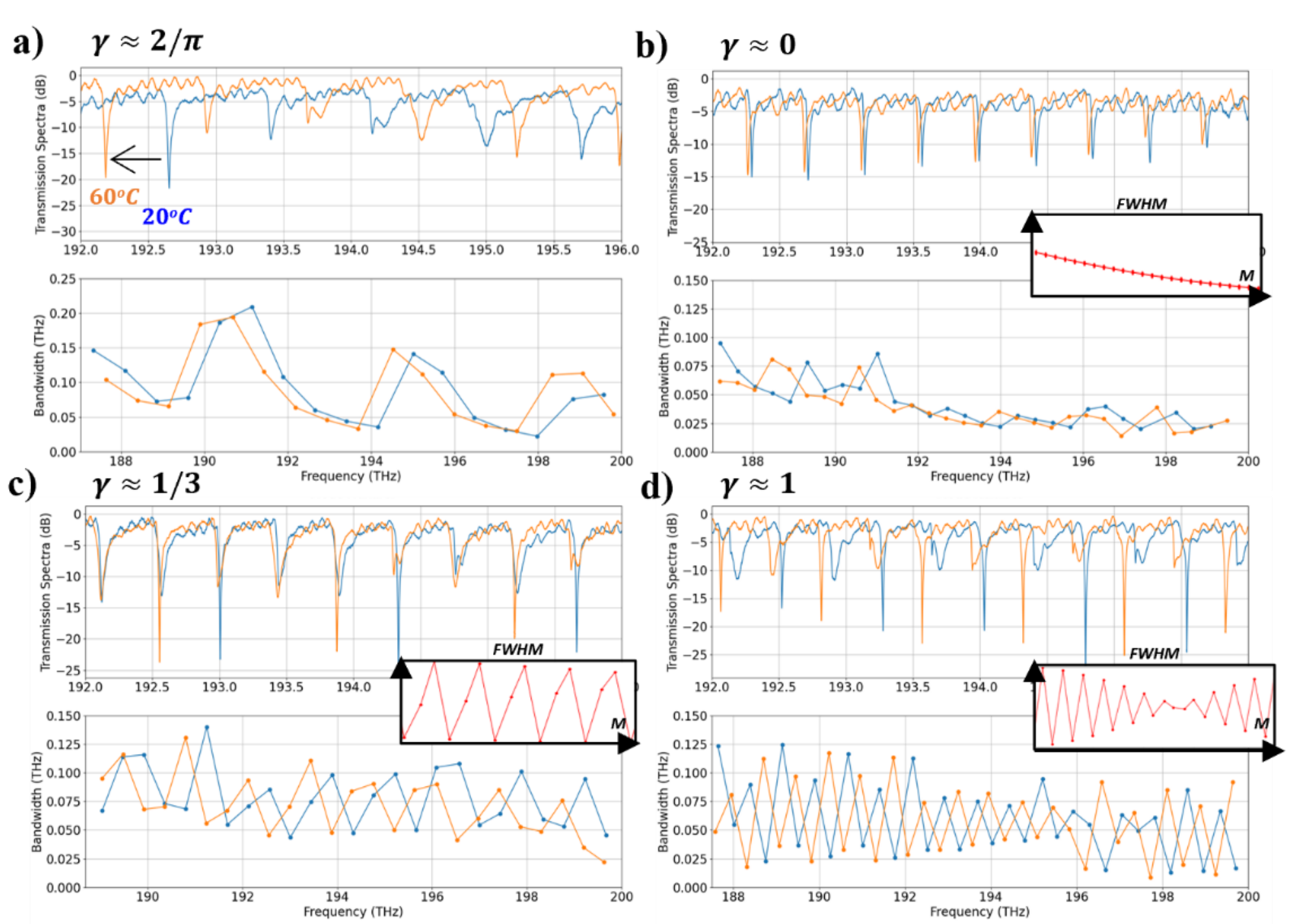

**Fig. S5. Temperature Sensitivity Measurements.** Measured Transmission Spectra and Bandwidth trends for a Theta Cavity with a) Diameter D = 28.5um, $\gamma \approx \frac{2}{\pi}$, b) $\gamma \approx 0$, c) $\gamma \approx \frac{1}{3}$, and d) $\gamma \approx 1$ at $\Delta T = 0$ (blue) and $\Delta T = 40^{\circ}C$. (orange) $\Delta\omega \approx 0.473$ THz per $\Delta T = 40^{\circ}C$.

`

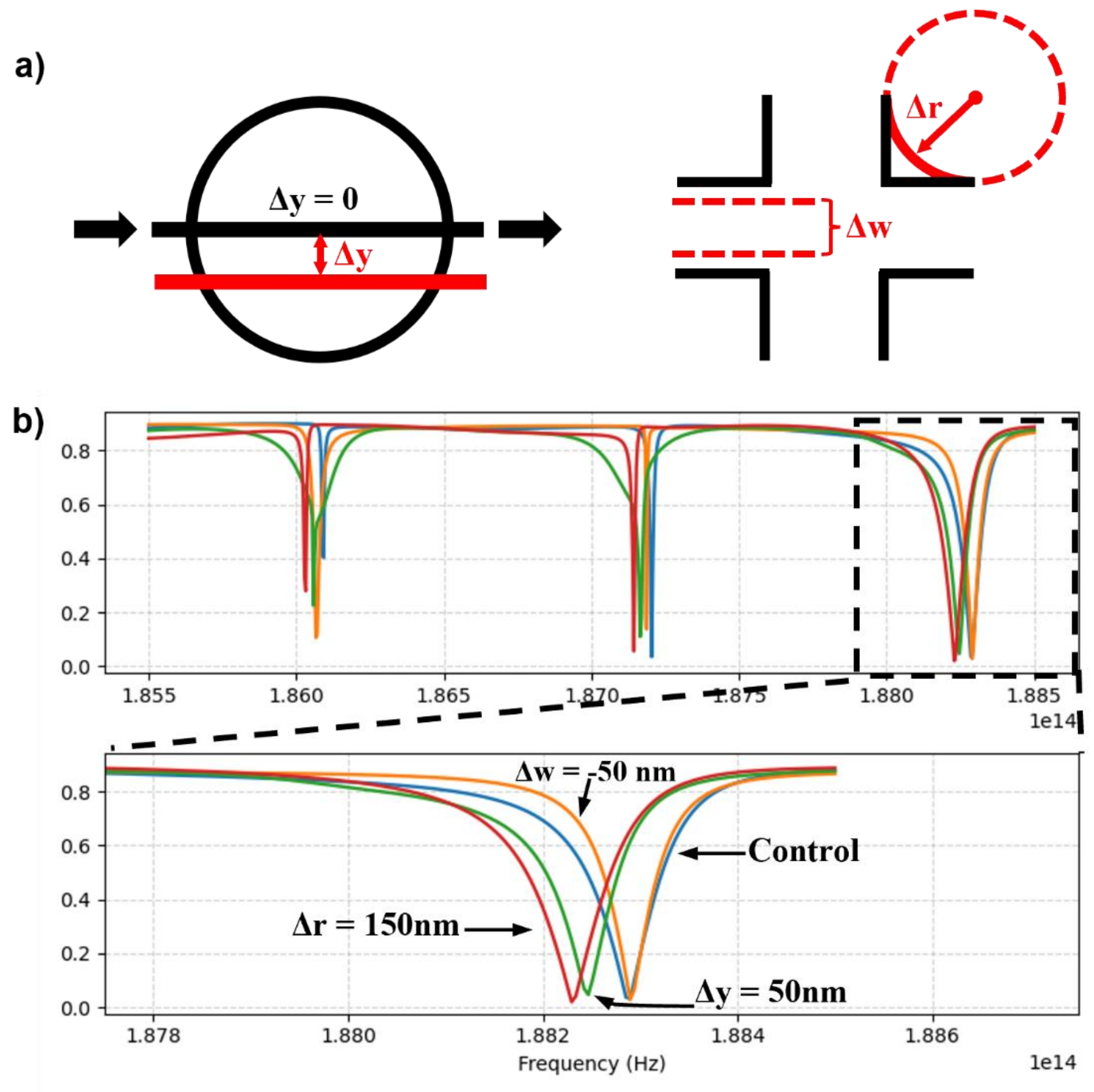

**Fig. S6. Simulation of Non-Idealities.** a) Schematic of a Theta Cavity visualizing non-idealities from a waveguide offset (Δy), change in waveguide width (Δw), and radius of curvature (Δr) of corners within the MSC junction. b) Simulated transmission spectra of Theta Cavities with Δw = -50nm , Δy = 50nm, Δr = 150nm, respectively.

`

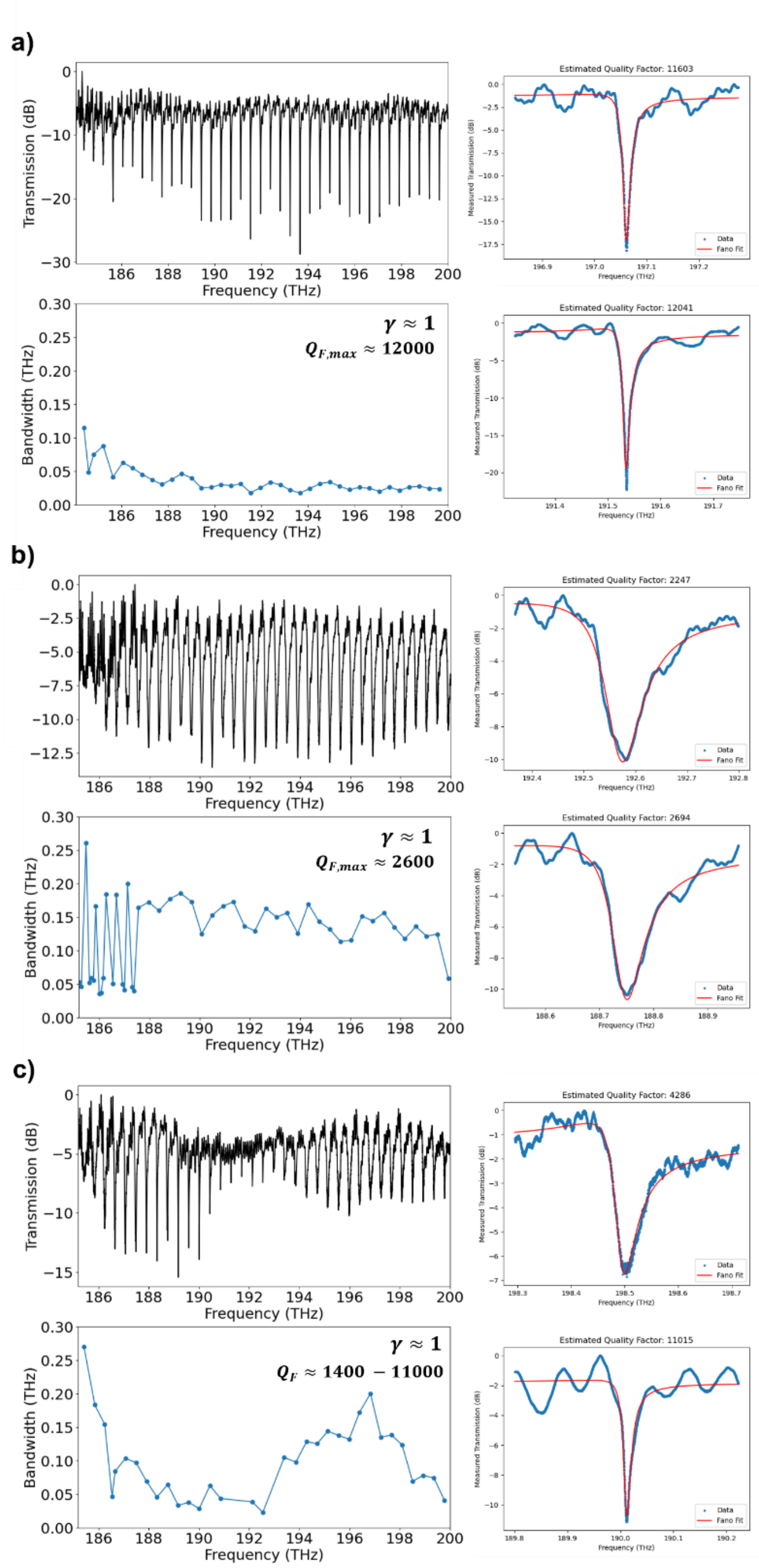

**Fig. S7. Measured Balanced Theta Cavity Spectral Analysis.** Measured transmission and bandwidth of Two geometrically identical balanced Theta Cavities showing monotonic spectral patterns with distinctly different mode behavior, sharp modes (a), broad modes (b), and transition between sharp and broad modes (c), due to slight differences in gamma (γ) due to fabrication variability.

`

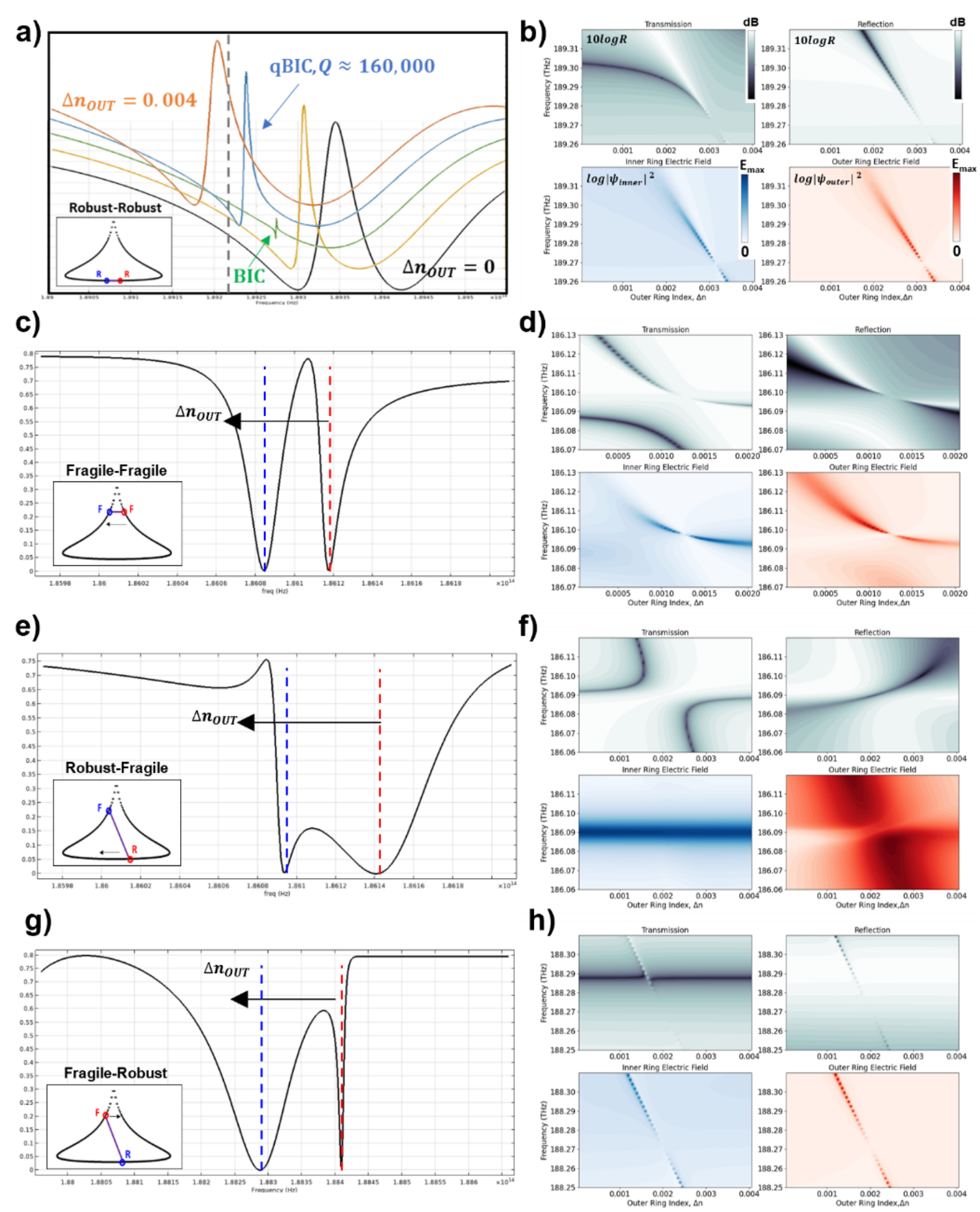

**Fig. S8. Simulated NTC Dispersion.** Simulation results of Nested Theta Cavities, where the parameters of the inner ring is held constant and the refractive index parameter of the outer ring is tuned ($\Delta n < 0.004$), to investigate mode hybridization between broad modes (Robust, R) and sharp modes (Fragile, F). Results are shown for representative coupled resonances Robust-Robust (a, b), Fragile-Fragile (c, d), Robust-Fragile (e, f), and Fragile-Robust (g, h), corresponding to the mode characteristics in the outer and inner rings, respectively. a, c, e, f) Simulated transmission spectra for different index offsets, overlayed with artificial offsets to show the progression of the spectral line shapes resulting from strong coupling. Insets show phase transition diagrams to illustrate the nature of the mode coupling. b, d, f, h) Contour maps plotting simulated transmission spectra, reflection spectra, and electric field enhancement in the inner and outer rings (z-axis) are plotted against the index offset (x-axis) and frequency (y-axis). Contour maps of the transmission spectra serve as an optical dispersion relation, revealing band hybridization dynamics. Contour maps of the reflection spectra show the progression of the spectral shape, revealing Fano-like modes and qBIC to BIC transitions. Contour maps of electric field enhancement for both the outer and inner rings reveal inner ring modes (blue), outer ring modes (red).

`

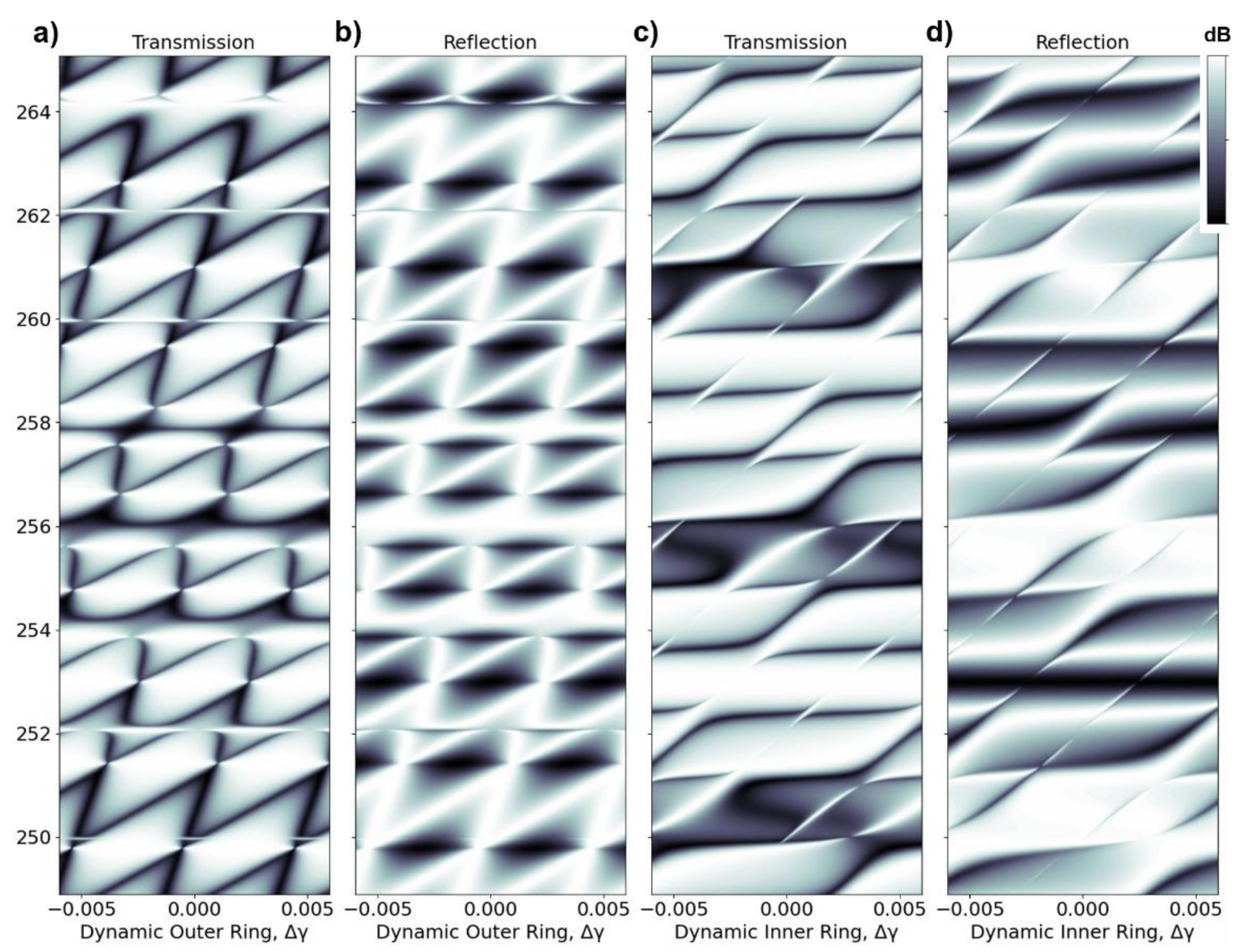

**Fig. S9. Band Hybridization in Transmission Dispersion.** Analytical results of Nested Theta Cavities, where the parameters of the outer (a, b) and outer (c, d) rings are individually tuned while the other is held constant. Contour maps plotting transmission (a, c) and reflection (b, d) are plotted against the gamma offset (x-axis) and mode number (y-axis), revealing correlations between bandgap size and relative phase between rings.

`

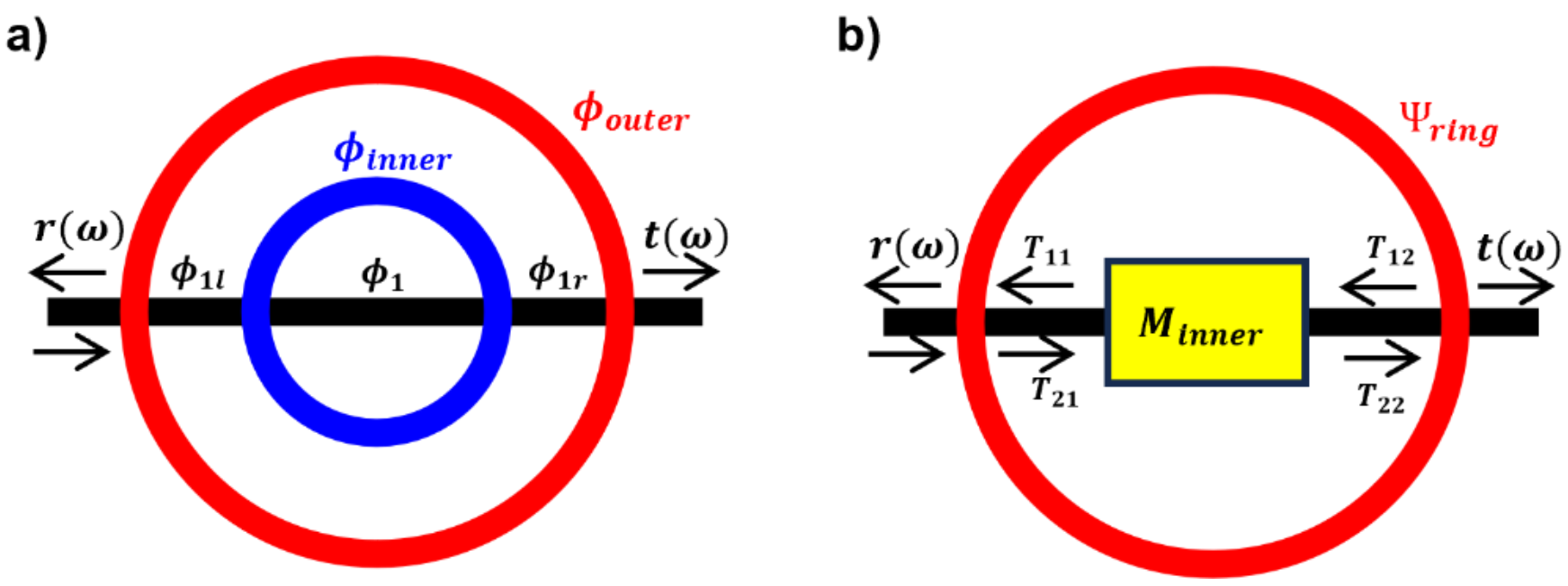

**Fig S10: Nested Theta Cavity Model.** (a) A simple illustration showing the physical model used to describe the wavefunctions in the dual, nested ring cavity. (b) An illustration showing the transfer matrix method used to solve the analytical model.

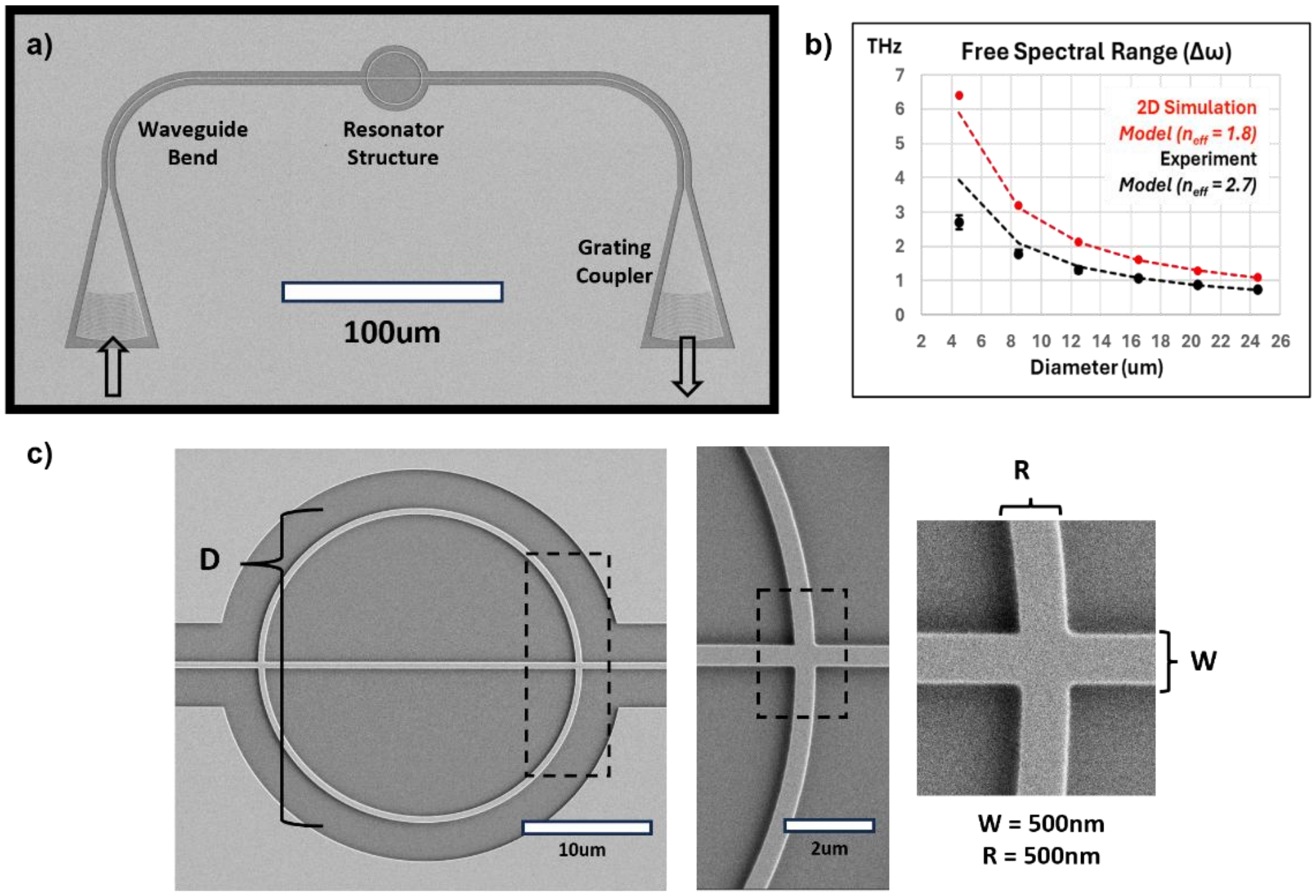

**Fig. S11. Device Fabrication Effects.** (a) Scanning Electron Microscopy image of a typical Circular Theta Cavity with grating couplers on SOI substrate. (b) Calculated free spectral range of SOI waveguides used for 2D simulation and experimental fabrication as a function of Theta Cavity diameter, the effective index is discerned by fitting the model to the free spectral range shown in inset. (c) Scanning Electron Microscopy image of circular MC junction where D = diameter, R = ring width, and W = center waveguide width

`

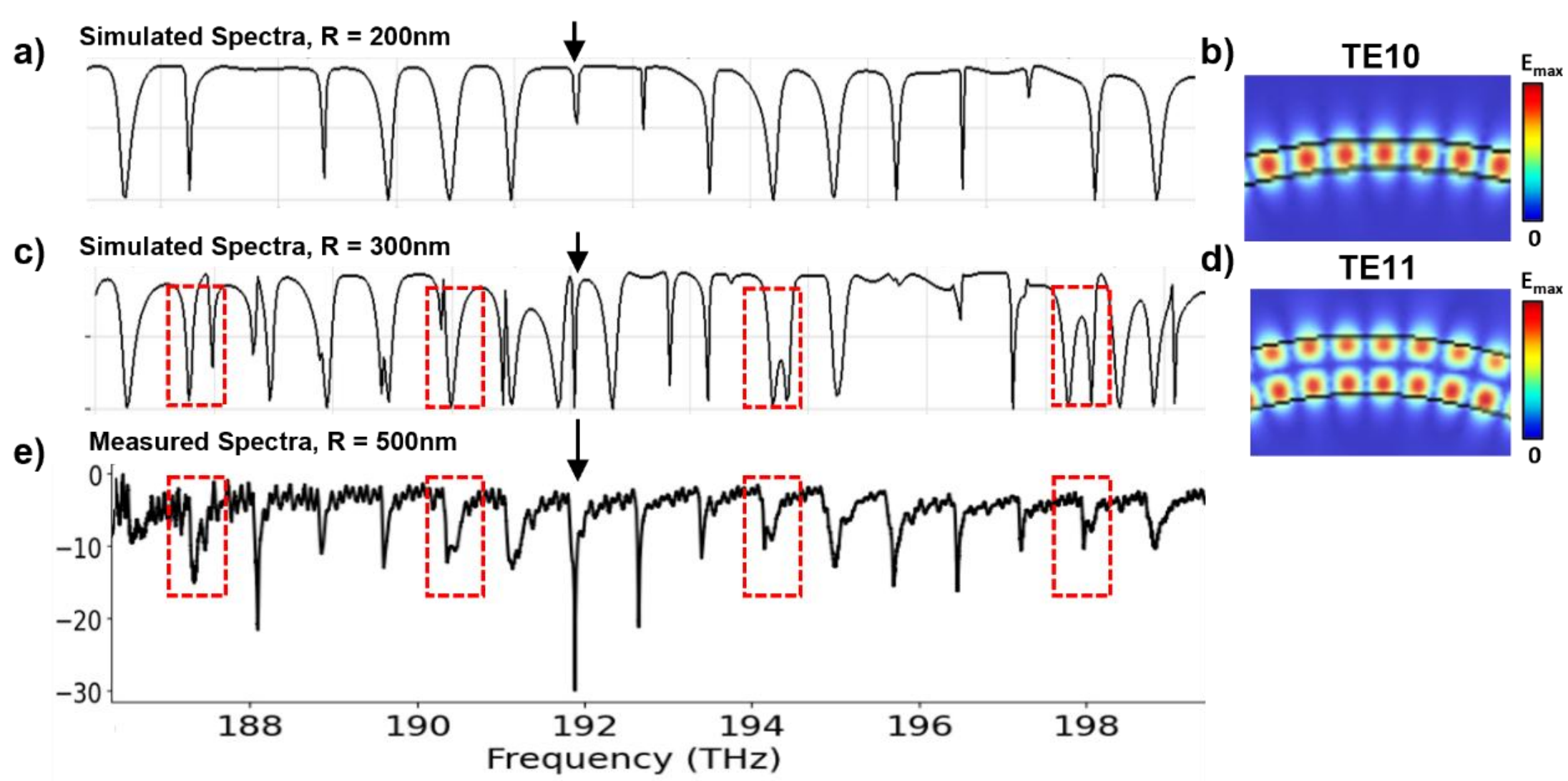

**Fig. S12. Multimode Waveguide Effects.** Simulated and measured transmission spectra of a Circular Theta Cavity (D = 25um) reveals signatures of multimode effects. The simulated spectra with waveguide width of R = 200 nm (a) and a map of the ring's electric field (b) shows the fundamental (TE10) mode is supported. The simulated spectra with waveguide width of R = 300 nm (c) and a map of the ring's electric field (d) shows the presence of higher order modes (TE11). e) Measured transmission spectra of a Circular Theta Cavity with ring waveguide width R = 500 nm, reveal line shapes that supports the presence of multimodes (highlighted by red boxes).

`

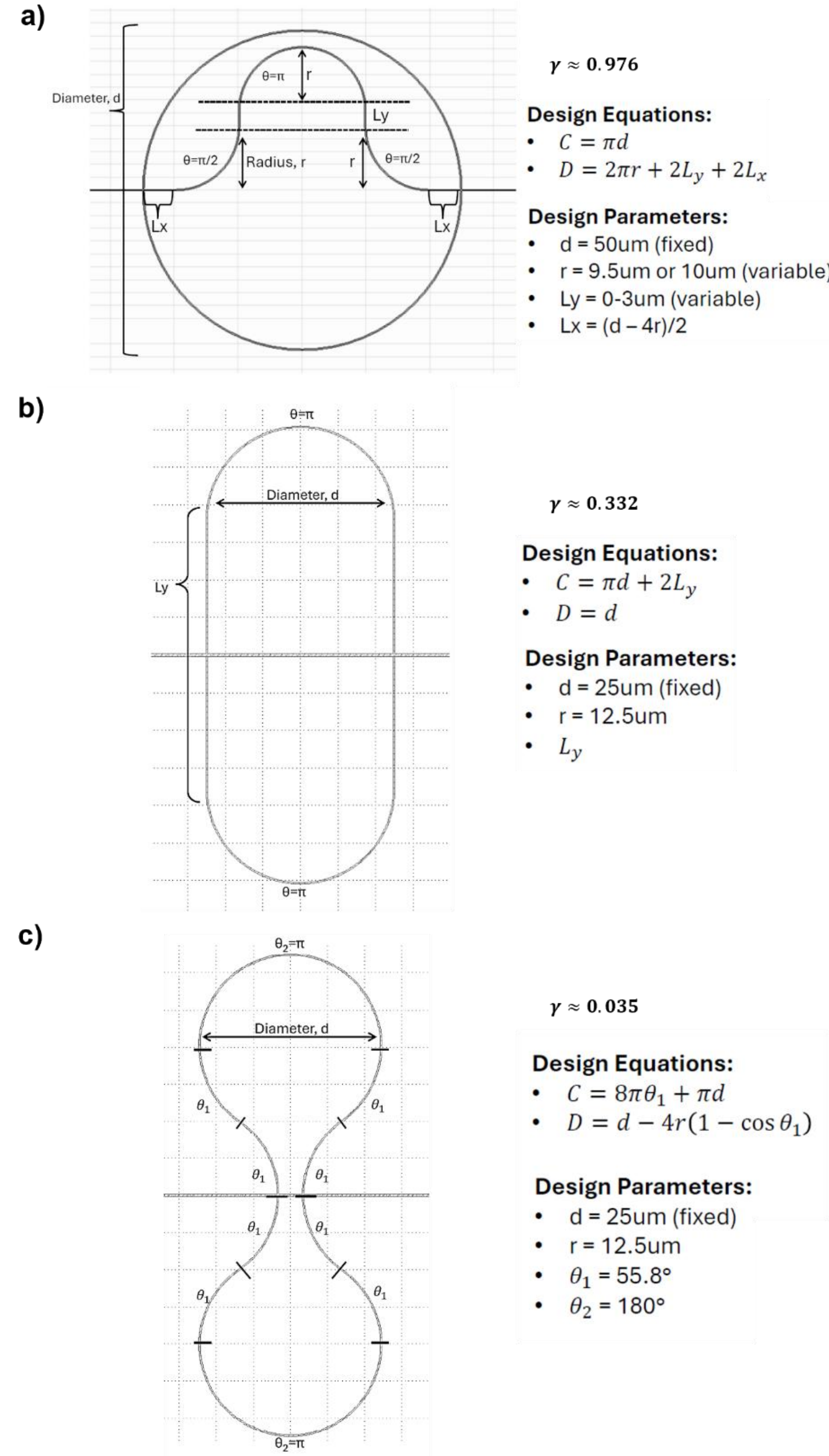

**Fig. S13. Silicon-on-Insulator Theta Cavity Design Limitations.** Design Schematics of Theta Cavities with (a) $\gamma \approx 0.976 \approx 0$, (b) $\gamma \approx 0.332 \approx 1/3$, and (c) $\gamma \approx 0.035$, showing design equations and geometric parameters.

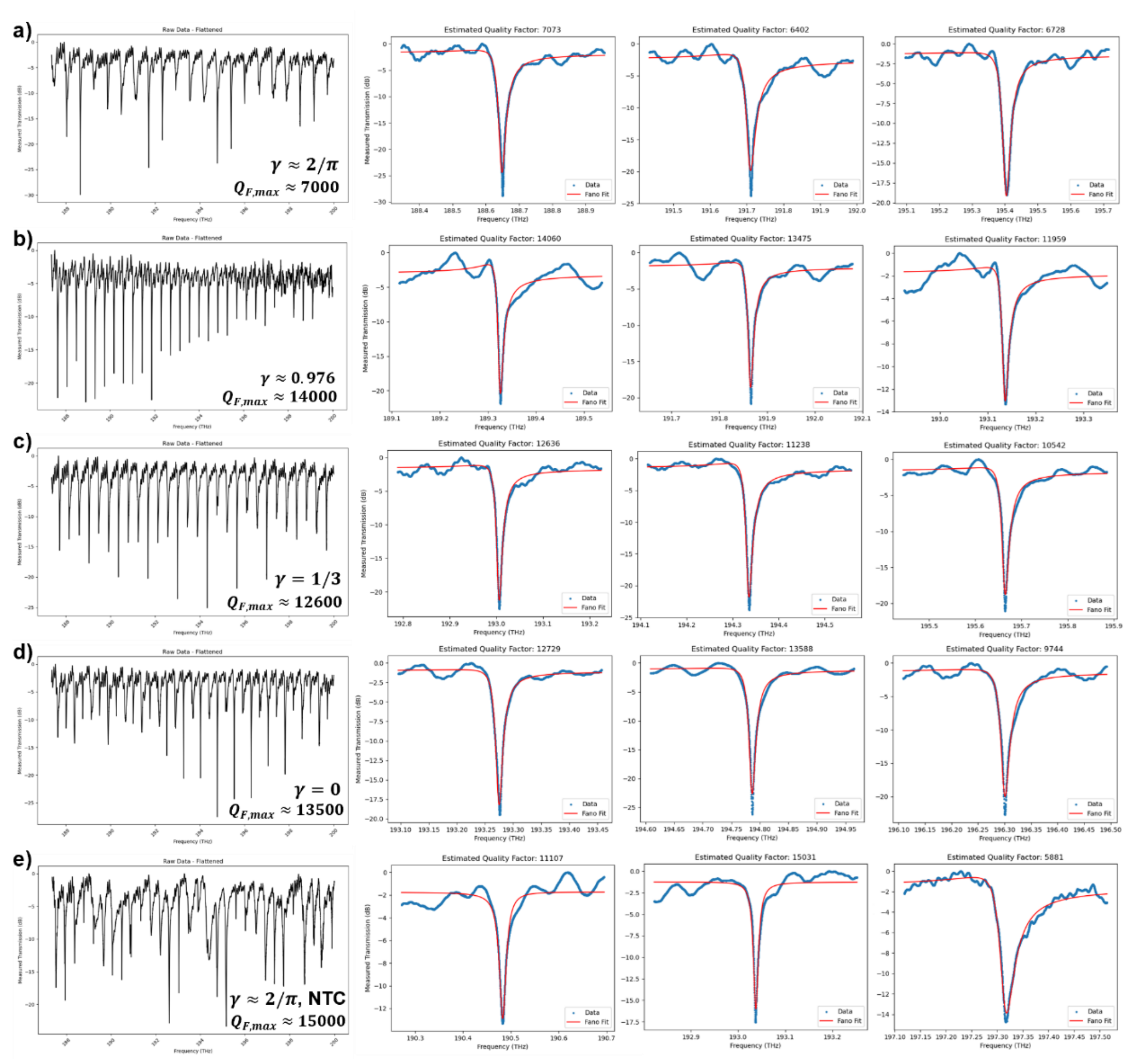

**Fig. S14. Quality Factor Extraction.** A plot of flattened measured spectra (left) and the fitted Fano line shape (red) is overlapped on the raw data (blue) for the highest quality factor modes in the spectra (right), the estimated quality factor is shown in the plot title. The quality factor extraction analysis was performed on spectra from various Theta Cavity devices including the Circular TC with D = 35 um (a), the balanced TC with γ ≈ 0.976 (b), the periodic TC with γ ≈ 0.332 (c), the oscillating with γ ≈ 0.035 (d), and the nested TC with $D_{inner}$ = 25um and $D_{outer}$ = 50nm (e).